\documentclass[12pt]{article}
\usepackage{tikz}

\usepackage{url}
\usepackage{amsmath}
\usepackage{amsfonts}
\usepackage{amssymb}
\usepackage{todonotes}
\definecolor{sangre}{rgb}{0.6,0.18,0.19}
\definecolor{dullmagenta}{rgb}{0.4,0,0.4}
\definecolor{darkblue}{rgb}{0,0,0.6}

\usepackage{relsize,etoolbox}
\usepackage[para,online,flushleft]{threeparttable}

\usepackage{amsmath}

\usepackage{bbm}

\usepackage{color, colortbl}
\definecolor{Gray}{gray}{0.95}
\usepackage{booktabs}
\usepackage{pdflscape}
\usepackage{amsfonts}
\usepackage{graphicx}
\usepackage{subcaption}
\usepackage{multirow}
\usepackage{epstopdf}
\usepackage{adjustbox}
\usepackage[titletoc,title]{appendix}

\definecolor{lavander}{cmyk}{0,0.48,0,0}
\definecolor{violet}{cmyk}{0.79,0.88,0,0}
\definecolor{burntorange}{cmyk}{0,0.52,1,0}

\def\lav{lavander!90}
\def\oran{orange!30}

\tikzstyle{peers}=[draw,circle,violet,bottom color=\lav,
                  top color= white, text=violet,minimum width=10pt]
\tikzstyle{superpeers}=[draw,circle,burntorange, left color=\oran,
                       text=violet,minimum width=20pt]
\tikzstyle{legendsp}=[rectangle, draw, burntorange, rounded corners,
                     thin,bottom color=\oran, top color=white,
                     text=burntorange, minimum width=2.5cm]
\tikzstyle{legendp}=[rectangle, draw, violet, rounded corners, thin,
                     bottom color=\lav, top color= white,
                     text= violet, minimum width= 2.5cm]
\tikzstyle{legend_general}=[rectangle, rounded corners, thin,
                           burntorange, fill= white, draw, text=violet,
                           minimum width=2.5cm, minimum height=0.8cm]

\usepackage[most]{tcolorbox}
 \tcbset{colback=white}

\usepackage[square]{natbib}
\usepackage{multibib}
\newcites{main}{References}

\usepackage{setspace}
\usepackage[margin=1in]{geometry}

\ifx\pdfoutput\undefined
\usepackage[backref=page,hypertex,colorlinks,urlcolor = darkblue,citecolor = sangre,linkcolor=blue]{hyperref}
\else
\usepackage[backref=page,pdftex,hypertexnames=false,colorlinks,urlcolor = darkblue,citecolor =  sangre,linkcolor=blue]{hyperref}
\fi
\usepackage[capitalize]{cleveref}
\crefname{appsec}{Section}{Sections}
\def\@mb@citenamelist{cite,citep,citet,citealp,citealt}
\begin{document}

\title{Coordinated Capacity Reductions and Public Communication in the Airline Industry%
	\thanks{This paper was previously circulated under the title ``Public Communication and Collusion in the Airline Industry."
	 We thank Yu Awaya, David Byrne,  Karim Chalak, Marco Cosconati, Kenneth G. Elzinga, Leslie Marx, Robert Porter, Mar Reguant, D. Daniel Sokol, and the seminar/conference participants at the DOJ, University of Florida, UVa, the 16th IIOC, 2018 BFI Media and Communication Conference, 2018 DC IO Day, NBER IO SI 2018, EARIE 2018, 2018 FTC Microeconomics Conference, 2018 PSU-Cornell Conference, 8th EIEF-UNIBO-IGIER Bocconi IO Workshop, and 2020 Next Generation of Antitrust Scholars Conference for their constructive feedback. We also thank the Buckner W. Clay Dean of A\&S and the VP for Research at UVa for financial support, and Divya Menon for outstanding research assistance. Aryal and Ciliberto acknowledge the Bankard Fund for Political Economy at the University of Virginia for support. Finally, we thank Aureo de Paula and three anonymous reviewers for their helpful feedback.}}

\author{Gaurab Aryal\thanks{ Department of Economics, University of Virginia, \href{mailto:aryalg@virginia.edu}{ aryalg@virginia.edu}.},
 Federico Ciliberto\thanks{Department of Economics, University of Virginia; 
 CEPR, London; DIW, Berlin,
 \href{mailto:ciliberto@virginia.edu}{ ciliberto@virginia.edu}.}, and
 Benjamin T. Leyden\thanks{ Dyson School of Applied Economics and Management, Cornell University; CESifo \href{mailto:leyden@cornell.edu}{leyden@cornell.edu}.}
}

\date{July 25, 2021}

\maketitle

\begin{abstract}
We investigate the allegation that legacy U.S. airlines communicated via earnings calls to coordinate with other legacy airlines in offering fewer seats on competitive routes. To this end, we first use text analytics to build a novel dataset on communication among airlines about their capacity choices. Estimates from our preferred specification show that the number of offered seats is 2\% lower when all legacy airlines in a market discuss the concept of ``capacity discipline.'' We verify that this reduction  materializes only when legacy airlines communicate concurrently, and that it cannot be explained by other possibilities, including that airlines are simply announcing to investors their unilateral plans to reduce capacity, and then following through on those announcements. \\
  (JEL: D22, L13, L41, L93)
\end{abstract}

\newpage
%\linenumbers

\section{Introduction}
There are two legal paradigms in most OECD countries meant to promote market efficiency, but that are potentially at odds with one another. On the one hand, antitrust laws forbid firms from communicating their strategic choices with each other to deter collusion. On the other hand, financial regulations promote open and transparent communication between publicly traded firms and their investors. While these latter regulations are intended to level the playing field among investors, policymakers have raised concerns that they may also facilitate anticompetitive behaviors. For example, the OECD Competition Committee notes that there are pro-competitive benefits from increased transparency, but ``information exchanges can ... offer firms points of coordination or focal points,'' while also ``allow[ing] firms to monitor adherence to the collusive arrangement'' \citepmain{OCED2011}. Thus, firms can be transparent about their future strategies in their public communications to investors---e.g., by announcing their intention to rein capacity---which can foster coordination among firms in offering fewer seats.\footnote{  Similar situations, where one set of laws is at odds with another, generating unanticipated consequences, often as antitrust violations, occur in many industries. For example, in the U.S. pharmaceutical industry, the tension between the FDA laws and patent law led to the Drug Price Competition and Patent Term Restoration Act (colloquially known as the Hatch-Waxman Act). This Act aims to reduce entry barriers for generic drugs, but it incentivized incumbent firms to ``Pay-for-Delay'' of generic drugs and stifle competition. For more, see \citemain{FedlmanFrondorf2017}. Other cases include \citemain{ByrneRoos2017} who document that gasoline retailers in Australia used a price transparency program called \emph{Fuelwatch} to initiate and sustain collusion. Furthermore, \citemain{BourveauSheZaldokas2019} document that with the increase in cartel enforcement, firms in the U.S. start sharing more detailed information in their financial disclosure about their customers, contracts, and products, which may allow tacit coordination in product markets.}

In this paper, we contribute to this overarching research and policy issue by investigating whether the data are consistent with the hypothesis that top managers of legacy U.S. airlines used their quarterly earnings calls to communicate with other legacy airlines to coordinate in reducing the number of seats offered.\footnote{An earnings call is a teleconference in which a publicly-traded company discusses its performance and future expectations with financial analysts and news reporters. Legacy carriers are Alaska Airlines (AS), American Airlines (AA), Continental Airlines (CO), Delta Airlines (DL), Northwest Airlines (NW), United Airlines (UA) and US Airways (US), and the low-cost carriers (LCC) are AirTran Airways (FL), JetBlue (B6), Southwest (WN) and Spirit Airlines (NK).} Specifically, we investigate whether legacy airlines used keywords associated with the notion of ``capacity discipline'' in their earnings calls to communicate to their counterparts their willingness to reduce offered seats in markets where they compete head-to-head.\footnote{  This idea that ``capacity discipline'' is used by airlines to signal their alleged intention to restrict supply has been applied in class-action lawsuits filed against a few airlines. \citemain{NYT2} and \citemain{NYT1} provide coverage of this concept in the popular press.
 See \citemain{RosenfieldCarltonGertner1997} and \citemain{Kaplow2013} for antitrust issues related to communication among competing firms.}

The airline industry is an appropriate industry to investigate the relationship between communication and coordinated reduction in capacities because it is characterized by stochastic demand, and \emph{private} and \emph{noisy} monitoring, both of which make coordination difficult without communication.\footnote{There is a precedence of accusation against the airlines for using communication technologies to coordinate. For example, in 1992, the U.S. DOJ alleged that airlines used the Airline Tariff Publishing Company's electronic fare system to communicate and collude, see, for example,  \citemain{Borenstein2004} and \citemain{Miller2010}. } Demand is stochastic, not least because of exogenous local events, such as the weather, unforeseen events at the airport, and spillovers from other airports.
Monitoring is private and noisy because, first, airlines do not instantaneously observe others' actions; second, they use connecting passengers to manage their load factors; and third, they observe only each other's list prices, not transaction prices. 

Recently, \citemain{AwayaKrishna2016}, \citemain{AwayaKrishna2017} and \citemain{Spector2015} have shown that firms may be able to use cheap talk--unverifiable and non-binding communication--to sustain collusion in environments with private and noisy monitoring, where collusion is otherwise unsustainable.\footnote{  There is a vast literature on market conduct and the behavior of cartels; see \citemain{Harrington2006,MailathSamuelson2006, HarringtonSkrzypacz2011}, and \citemain{MarshallMarx2014}. Among others, \citemain{Porter1983, GreenPorter1984} study collusion under imperfect monitoring where all firms observe the same (noisy signal) price. In their setting, access to communication technology does not have any effect because the profits from public perfect equilibrium with or without communication are the same. 
Some examples where communication helped collusion are \citemain{GenesoveMullin2001}, \citemain{Wang2008,Wang2009}, \citemain{ClarkHoude2014}, and \citemain{ByrneRoos2017}, among others. 
}
In our context, airlines have access to public communication technology, their quarterly earnings calls, through which they can \emph{simultaneously} communicate with other airlines.\footnote{ \label{footnote:othertalk} Airlines may have other avenues for coordination, e.g., via industry conferences and trade organization events  \citepmain{AwayaKrishna2020} and common-ownership \citepmain{AzarSchmalzTecu2018}. However, quarterly earnings call are ideal for our purpose because they occur at regular intervals, \emph{every} publicly listed airline uses them, and we observe the conversation. Our decision to consider only communication through earnings calls can be viewed as \emph{conservative} because we cannot account for any amount of relevant communication outside this medium and underestimate the negative relationship between communication and capacity. And lastly, we focus only on simultaneous messaging among (legacy) airlines and do not distinguish intra-quarter timing of airlines because determining if there is a ``leader'' among the airlines by following, say, \citemain{ByrneRoos2017}, would require higher-frequency (e.g., daily) data on communication.}
 
We build an original and novel dataset on the content of public communication from earnings calls to measure communication and assess its relationship with capacity. The Securities and Exchange Commission (SEC) requires all publicly traded companies in the U.S. to file a quarterly report, which is accompanied by an earnings call---a public conference call where top executives discuss the report's content with analysts and financial journalists. We collected transcripts of all such calls for 11 airlines from 2002:Q4 to 2016:Q4. We then classified each earnings call as either pertinent or not pertinent, depending on whether the executives on the call declared their intention of engaging in capacity-discipline or not.\footnote{  Other papers that use ``\emph{text as data}" \citepmain{GentzkowKellyTaddy2017}, include \citemain{Leyden2017}, who uses text descriptions of smartphone and tablet \emph{apps} to define relevant markets, \citemain{GentzkowShapiro2010}, who use
phrases from the \emph{Congressional Record} to measure the slant of news media, and \citemain{HobergPhilips2016}, who use the text descriptions of businesses included in financial filings to define markets.}

We estimate the relationship between communication and carriers' monthly, market-level capacity decisions using the Bureau of Transportation Statistics's T-100 Domestic Segment dataset, which contains domestic non-stop segment data reported by both U.S. and foreign air carriers. To that end, we regress the log of seats offered by an airline in a market in a month on an indicator of whether \emph{all} legacy carriers operating in that market discuss capacity discipline. Given that airlines' capacity decisions depend on a wide variety of market-specific and overall economic conditions, we include covariates to control for such variation across markets and carriers over time.

We find that when all legacy carriers operating in an airport-pair market communicate about capacity discipline, the average number of seats offered in that market is 2.02\%  lower. To put this number in perspective, we note that the average change in capacity among legacy carriers in comparable markets where communication does not occur is 3.67\%. So a 2.02\% reduction in capacity associated with the phrase ``capacity discipline'' accounts for more than 50\% of this average change, which is economically significant.

Capacity reductions have the potential to benefit consumers if they lead to a more optimal scheduling of flight departure times at the airports without affecting ticket fares. However, we (i) do not find evidence to support the hypothesis that carriers adjust the crowding of departures, and, furthermore, we (ii) find that communication is positively associated with fares.  
So, even though we do not estimate the social value of communication \citepmain{MyattWallace2015}, our estimates suggest that the carriers' capacity reductions are economically significant and they most likely harm consumers.

Nonetheless, we face two primary identification challenges in investigating the accusation that legacy U.S. carriers are using their earnings calls to coordinate capacity reduction. First, there may be a more straightforward, alternative explanation for our findings. In particular, it might be that airline executives are communicating to investors their intention to reduce capacity, not because they want to coordinate, but because reducing capacity is the best response to negative demand forecasts. In other words, our results may be evidence that earnings calls are serving their ostensible purpose. 

We address this concern in three ways. 
First, we find that legacy carriers unilaterally discussing capacity discipline is not associated with them reducing capacities. Second, we find that the capacity is not lower in monopoly markets when legacy carriers discuss capacity discipline. Finally, we find that legacy carriers do not decrease their capacity when all but one of the legacy carriers serving a market have discussed capacity discipline. Suppose discussions of capacity discipline were meant to inform investors about the carrier's future actions. In that case, we should see a reduction in all three of these cases. 

Second, an airline could be using earnings calls to truthfully share its payoff relevant private information with other airlines, which, when others do the same, could induce correlation in their capacity plans. Importantly, this alternative explanation does not require airlines to actively coordinate their capacity choices, as long as they communicate truthfully. 

We do not believe that this explains our findings. First, we note that \citemain{Clarke1983}, \citemain{Galor1985}, and \citemain{Li1985} have shown that firms do \emph{not} have an incentive to share their payoff relevant private information about demand with others unless they intend to coordinate on an action, e.g., capacity choice. Second, if this hypothesis is correct, then it implies that the likelihood of us observing a reduction in capacity by an airline would increase with the number of legacy airlines communicating, irrespective of the said airline's private information. If airlines were only sharing their information, then an airline should be responsive to others' announcements. We show that, contrary to this information-sharing hypothesis, even when all of a legacy carrier's legacy competitors in a market communicate, if the carrier itself does not communicate, then it does not reduce its capacity. However, this result is consistent with airlines using earnings calls to coordinate on their capacities. 

\section{Institutional Analysis and Data}

In this section we introduce the legal cases that motivate our analysis, explain how we use Natural Language Processing (NLP) techniques to quantify communication among airlines, present our data on the airline industry, and show that airlines have flexible capacity at the market level.

\subsection{Legal Case}\label{section:legal}

On July 1, 2015, the \emph{Washington Post} reported that the U.S. Department of Justice (DOJ) was
investigating possible collusion to limit available seats and maintain higher fares in U.S. domestic airline markets by American, Delta, Southwest, and United (Continental) \citepmain{Wapo2015}. It was also reported that the major carriers had received Civil Investigative Demands (CID) from the DOJ requesting copies, dating back to January 2010, of all communications the airlines had had with each other, Wall Street analysts, and major shareholders concerning their plans for seat capacity and any statements to restrict it. The CID requests were subsequently confirmed by the airlines in their quarterly reports.\footnote{ In Appendix \ref{section:20102015} we consider whether our results vary before and after the January, 2010 threshold, and the July, 2015 reporting of the DOJ investigation.}

Concurrently, several consumers filed lawsuits accusing American, Delta, Southwest, and United of fixing prices, which were later consolidated in a
multi-district litigation. The
case is currently being tried in the U.S. District Court for the District of Columbia.\footnote{   This
 is the ``Domestic Airline Travel Antitrust Litigation'' case, numbered 1:15-mc-01404 in the US District Court, DC.}
Another case, filed on August 24, 2015, in the U.S. District Court
of Minnesota against American, Delta, Southwest Airlines, and United/Continental, alleges that the
companies conspired to fix, raise, and maintain the price of domestic air travel services in violation
of Section 1 of the Sherman Antitrust Act.\footnote{  Case 0:15-cv-03358-PJS-TNL, filed
8/24/2015 in the US District Court, District of Minnesota. In November 2015, this case was
transferred to the District Court in DC. At the time of this writing, American Airlines and Southwest have settled the class action lawsuits.}

The lawsuits allege that the airline carriers collusively impose ``capacity discipline'' in the form of limiting flights and seats \textit{despite increased demand and lower costs}, and that the four airlines implement and police the agreement through \textit{public signaling} of \textit{future} capacity decisions.\footnote{   The consumers' lawsuits also stress the role of financial analysts who participate at the quarterly earnings call. See \citemain{AzarSchmalzTecu2018} for a recent work on the role of institutional investors on market conduct. We instructed our research assistant (RA) to find all instances where institutional investors were the first to bring up capacity discipline. The RA found only three such instances. Therefore, we decided not to consider the role of institutional investors in our analysis.}  In particular, one of the consumers' lawsuits reported several statements made by the top managers of American, Delta, Southwest, United, and other airlines. The statements were made during quarterly earnings calls and various conferences.\footnote{  For example, during the US Airways 2012:Q1 earnings call, the CFO of US Airways Derrick Kerr said \begin{quote}
	``.. mainline passenger revenue were \$2.1 billion, up 11.4\% as a result of the strong pricing environment and continued industry capacity discipline.'' -- US Airways. \end{quote}
	and in the Delta's earnings calls for the same quarter Delta's CEO Richard Anderson said 
	\begin{quote}
	``You've heard us consistently state that we must be disciplined with capacity.'' -- Delta 
\end{quote}}

These lawsuits provide the foundation to build a vocabulary from the earnings calls that can capture legacy airlines' (alleged) intention to restrict their offered capacity. To that end, we have to consider both the semantics (airlines' intention to rein in capacity) and the syntax (which keywords are used) of the earnings call reports. 
Next, we explain the steps we take to measure communication.
   
\subsection{Earnings Call Text as Data}\label{section:ML}

All publicly traded companies in the U.S. are required to file a quarterly report with the
SEC. These reports are typically accompanied by an earnings
call, which is a publicly available conference call between the firm's top management and the analysts and reporters
covering the firm. Earnings calls begin with statements from some or all of the corporate
participants, followed by a question-and-answer session with the analysts on the call. Transcripts of calls are readily available, and we assume that carriers observe their competitors' calls.

\begin{figure}[t!]\caption{Transcript Availability}\label{fig:transcript_availability}\centering
\includegraphics[scale=0.5]{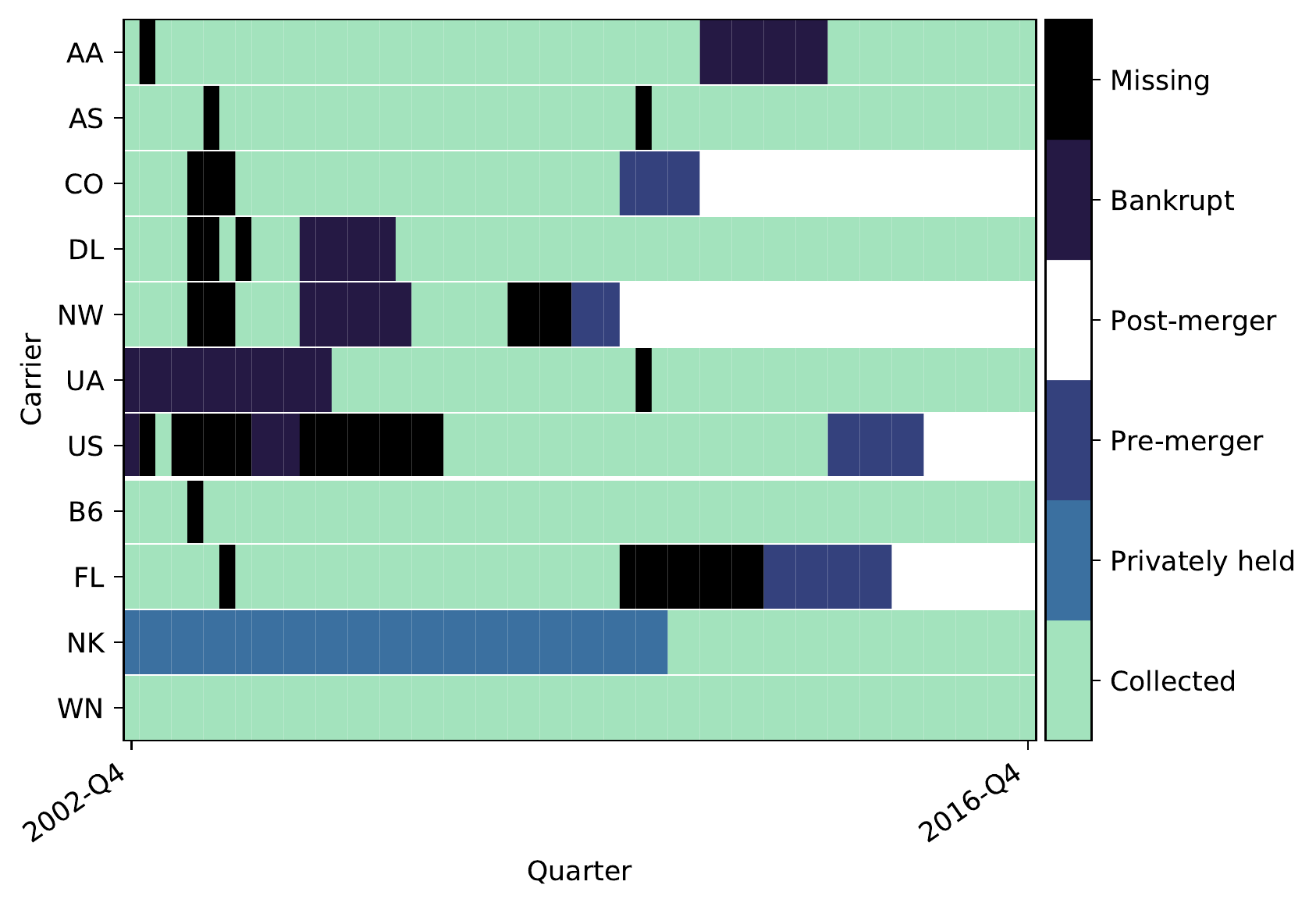} 
\caption*{\footnotesize Notes. This figure shows the availability or non-availability of transcripts for 11 airlines. The x-axis denotes the time  year and quarter, and the y-axis denote the name of the airline. Each color/shade denotes the status of the transcript.}
\end{figure}

We collected earnings call transcripts for 11 airlines, for all quarters from 2002:Q4 to 2016:Q4 from  LexisNexis (an online database service) and Seeking Alpha (an investment news website). Figure \ref{fig:transcript_availability} indicates the availability of transcripts in our sample for each of the 11 airlines. As the figure shows, transcripts are available for most of the quarters except under (i) Bankruptcy---five carriers entered bankruptcy at least once during the sample period; (ii) Mergers and acquisitions---airlines did not hold earnings calls in the interim period between the announcement of a merger and the full operation of the merger; (iii) Private airlines---Spirit Airlines, which was privately held until May 2011, neither submitted reports nor conducted earnings calls prior to its initial public offering; and (iv)  Other reasons---in a few instances the transcripts were unavailable for an unknown reason. In all cases where a call is unavailable, we assume the carrier cannot communicate to its competitors.\footnote{ \label{footnote:othertalk-conservative} Of course, the airlines may have other means to communicate, that we do not observe (e.g., see \cref{footnote:othertalk}).  To the extent to which airlines use other, unobserved, means of communications when earnings calls are unavailable our estimate will be biased toward zero (or positive).}

The key step of our empirical analysis is to codify the informational content in these
quarterly earnings calls into a dataset that can be used to see how capacity choices change over
time in response to communication among legacy carriers. Before delving into the conceptual challenges, we note two preliminary steps. Every statement made by the
operator of the call and the analysts is removed from the transcripts, as are common English ``stop words'' such as ``and'' and ``the.''
Then we tokenize (convert a body of text into a set of words or phrases) and lemmatize (reduce words to their dictionary form) the text from the earnings
calls. For example, the sentence ``The disciplined airline executive was discussing capacity discipline'' would be reduced to \{\texttt{discipline}, \texttt{airline}, \texttt{executive}, \texttt{discuss}, \texttt{capacity}, \texttt{discipline}\}.
 This process allows us to abstract from the
inflectional and derivationally related forms of words to better focus on the
substance/meaning of the transcripts.

To do so, we use a combination of NLP techniques and manual review to identify a list of words or phrases that are potentially indicative of managers communicating their intention to cooperate with others in restricting their capacity. Although in most cases managers specifically use the term ``capacity discipline," managers sometimes use other word combinations when discussing capacity discipline. This identification is a time-consuming process, and it is the focus of the remainder of this section.\footnote{ In \cref{sec:iv_cond_exog}, we also use NLP to identify words that can be used to evaluate conditional exogeneity in our setting.}

To codify the use of the phrase ``capacity discipline'' and other combinations of words that carry an analogous meaning, we begin by coding ``capacity discipline'' with a categorical variable $\texttt{Carrier-Capacity-Discipline}_{j,t}$, which takes the value $1$ if that phrase appears in the earnings call transcript of carrier $j$ in the year-quarter preceding the month $t$ and $0$ otherwise.

In many instances airline executives do not use the exact phrase ``capacity discipline,'' but the content of their statements is closely related to the notion of capacity discipline, as illustrated in the following text (emphasis added):
    \begin{quote} ``We intend to at least maintain our competitive position. And so, what's needed here, given fuel prices, \emph{is a proportionate reduction in capacity across all carriers in any given market}. And as we said in the prepared remarks, we're going to initiate some reductions and we're going to see what happens competitively. And if we find ourselves going backwards then we will be very capable of reversing those actions. So, this is a real fluid situation but clearly \emph{what has to happen across the industry is more reductions from where we are} given where fuel is running." -- Alaska Airlines, 2008:Q2.
	\end{quote}
\noindent Our view is that this instance and other similar ones should be interpreted as
conceptually analogous to uses of the phrase ``capacity discipline."
Yet in other cases it is arguable whether the content is conceptually analogous to the one of
``capacity discipline,'' even though the wording would suggest so. For example, consider the
following cases:

    \begin{quote}
    ``We are taking a disciplined approach to matching our plan capacity
    levels with anticipated levels of demand" -- American Airlines, 2017:Q3

    $ $

    ``We will remain disciplined in allocating our capacity in the markets that will generate the highest profitability." -- United Airlines, 2015:Q4
    \end{quote}

These statements, and others like these, cannot be easily categorized as a clear intention of the airlines to reduce capacity.\footnote{ Airlines can change the capacity across markets in multiple ways. They can remove an aircraft from a domestic market and keep it in a hangar, or they can move it to serve an international route, or they can reassign that plane to another domestic market. The airlines can also change the ``gauge'' of an aircraft, i.e., increase or decrease the number of seats or change the ratio of business to coach seats. Additionally, in markets where carriers outsource some flights and/or routes to regional carriers, moving capacity should be even easier. All of these options are discussed in conference calls.} On one hand, the ``anticipated levels of demand'' depend on the competitors' decisions, and thus one could interpret this statement as a signal to competitors to maintain capacity discipline. On the other hand, an airline should not put more capacity than what is demanded because that implies higher costs and lower profits.

We take a conservative approach and code all these instances as ones where the categorical
variable $\texttt{Carrier-Capacity-Discipline}_{j,t}$ is equal to 1. This approach is conservative because it assumes that the airlines are coordinating their strategic choices more often than their words would imply, and would work against finding a negative relation. In other words, we design our coding to err on the side of finding false negatives (failing to reject the null hypothesis that communication  is not correlated with a decrease in capacity), rather than erring on the side of finding false positives. We take this approach because our analysis includes variables that control for year, market, and year-quarter-carrier specific effects that control for many sources of unobserved heterogeneity that might explain a reduction of capacity driven by a softening of demand. Therefore, our coding approach makes us \textit{less} likely to find evidence of coordination even when airlines are coordinating.

In practice, to identify all the instances where the notion of capacity discipline was present but the phrase ``capacity discipline'' was not used, we used NLP to process all transcripts and flag those transcripts where the word ``capacity'' was used \textit{in conjunction with} either the word ``demand'' or ``GDP.'' This filter identified 248 transcripts, which we read manually to classify as either pertinent or not pertinent for capacity discipline. If the transcript was identified by all three of us as pertinent, then we set the variable $\texttt{Carrier-Capacity-Discipline}_{j,t}=1$, and zero otherwise. Out of the 248 transcripts, 105 contained statements that we deemed pertinent.\footnote{ Besides the coding approach described above, we had a research assistant independently code all transcripts, and coded all transcripts only using the automated, NLP approach. We discuss these approaches, and the results of estimating our primary model with these datasets, in Appendix \ref{appendix:ind_verification}.}

Table \ref{tab.transcript_statistics} presents the summary statistics of  
$\texttt{Carrier-Capacity-Discipline}_{j,t}$.
We have 320 earnings calls transcripts for the legacy carriers, and 40.9\% include content
associated with the notion of capacity discipline. We have fewer transcripts for LCCs, JetBlue and
Southwest, and content associated with capacity discipline is much less frequent. Overall, we have
520 transcripts and $\texttt{Carrier-Capacity-Discipline}_{j,t}=1$ in $29.2\%$ of
them.
Table \ref{tab.transcript_statistics} suggests that the LCCs, including
Southwest (WN), are much less likely to talk publicly about capacity discipline. 
In view of this data feature, in our empirical
exercise, we focus only on communication by legacy carriers.

\begin{table} \caption{Frequency of Communication}\label{tab.transcript_statistics}\centering
\begin{tabular}{llll}\toprule
	&Mean&SD&N \\
\midrule
\textbf{Carrier Type}&&& \\
Legacy&0.409&0.492&320 \\
LCC&0.124&0.331&89 \\
Jet Blue&0.109&0.315&55 \\
Southwest&0.071&0.260&56 \\
\midrule
\textbf{Total}&0.292&0.455&520 \\
\bottomrule
\end{tabular}
\caption*{\footnotesize Notes. Fraction of earnings calls where \texttt{Carrier-Capacity-Discipline}
is equal to one.} \end{table}

\subsection{Airline Data}\label{section:data}
We use three datasets for the airline industry: the Bureau of Transportation Statistics's (BTS) T-100 Domestic Segment for U.S. carriers,  the BTS's Airline On-Time Performance database, and a selected sample from the OAG Market Intelligence-Schedules dataset. We consider the months between 2003:Q1 and 2016:Q3 (inclusive). The BTS's T-100 Domestic Segment for U.S. carriers contains domestic non-stop segment (i.e., route) data reported by U.S. carriers, including the \textit{operating} carrier, origin, destination, available capacity, and load factor. The BTS's Airline On-Time Performance database reports flight times.

In many instances, regional carriers, such as SkyWest or PSA, also operate on behalf
of the \textit{ticketing} carriers. The regional carriers might be subsidiaries fully owned
by the national airlines, e.g., Piedmont, which is owned by American (and prior to that by U.S.
Airways), or they might operate independently but contract with one or more national carrier(s), e.g., SkyWest. 
To allocate capacity to the appropriate \textit{ticketing} carriers, we merge our data with the data from the OAG Market Intelligence, which contains information about the operating and the ticketing carrier for each segment at the quarterly level. Using this merged dataset, we allocate the available capacity in each route in the U.S. to the ticketing carriers, which are the carriers of interest.\footnote{A ticketing carrier is considered to have served a given market in a given month if it performed at least four flights in that month. We aggregate a set of particularly small ticketing carriers into a single ``Fringe'' carrier in our data.} We consider only routes between airports located in the proximity of a Metropolitan Statistical Area in the U.S.\footnote{ We use the U.S. DOC's 2012 data to identify Metropolitan Statistical Areas in the U.S. We also perform the empirical analysis where markets are defined by the origin and destination cities, rather than airports in Appendix \ref{sec:city-pair}.}

\subsection{Alignment of Earnings Calls and Airline Capacity}\label{sec:data_merge}

We investigate the relationship between communication via earnings calls and capacity decisions in the quarter following an earnings call --- i.e., in the intervening time between earnings calls. Earnings calls typically take place in the middle of the first month following a quarter. We use the content from a call, e.g., for Q1, occurring in mid-April, to define $\texttt{Carrier-Capacity-Discipline}_{j,t}$ for the months of May, June, and July.\footnote{ An alternative approach would be to associate the Q1 call taking place in mid-April with the capacity data for \emph{April}, May, and June. In Appendix \cref{section:realquarters} we present our primary results under this alternative approach. The results are similar to what we find under our preferred approach.}

We maintain that airlines can change route capacity (scheduled seats) by adding, removing, or changing flights, or the number of seats on a flight (up- or down-gauging), within few weeks of the scheduled departure day. We do not require that airlines regularly change their capacity in the days or weeks before takeoff across all market, but simply that \emph{they are able to make changes on relatively short notice in selected markets}. We have several pieces of evidence that support our timing assumption. First, Delta's ``Schedule Change and Ticket Revalidation Policy'' notes that ``airlines routinely change their flight schedules for a variety of reasons,'' such as ``seasonal demands, \dots, new routes, changes to ... operating times, [and] flights that no longer operate.'' Indeed, Delta further notes that, ``most schedule changes occur outside of 7 days before travel,'' which suggests a non-negligible number of changes occur within as little as a month before takeoff \citepmain{Delta2021}.  Second, there are court documents that provide additional evidence of airlines' abilities to make short-run changes. For example, the Memorandum and Order issued in the antitrust case \emph{United States of American v. AMR Corporation, American Airlines, Inc., and AMR Eagle Holding Corporation} documents several instances where airlines make the strategic decision to add or remove flights within days, and then enact those decisions within as little as two to three weeks \citepmain{MO2001}.

\subsection{Variable Definitions}\label{section:variabledefinitions}

We say that legacy airlines are communicating with each other when \textit{all} of the legacy airlines serving a market with at least two legacy carriers discuss capacity discipline. Letting $J_{m,t}^\texttt{Legacy}$ be the
set of legacy carriers in market $m$ at time $t$, we define a new variable,
\begin{eqnarray*}
&&\texttt{Capacity-Discip}\texttt{line}_{m,t}=\\
&&\,\begin{cases}
	\mathbbm{1}\left\{\texttt{Carrier-Capacity-Discipline}_{j,t}=1 \;\forall j\in J_{m,t}^{\texttt{Legacy}}\right\}& \!\!\!,\qquad|J_{m,t}^{\texttt{Legacy}}| \ge 2\\
	0&\!\!\! ,\qquad|J_{m,t}^{\texttt{Legacy}}| < 2\\
\end{cases}
\end{eqnarray*}
Thus, $\texttt{Capacity-Discipline}_{m,t}$ indicates whether all of the legacy carriers in $m$ discussed capacity discipline, conditional on two or more legacy carriers serving that market for that month.\footnote{  In \citemain{AwayaKrishna2016, AwayaKrishna2017} firms communicate simultaneously, and it is crucial for the construction of their equilibrium. For example, Awaya and Krishna write, ``The basic idea is that players can monitor each other not only by what they `see'---the signals---but also by what they `hear'---the messages that are exchanged'' \citemain[page 515]{AwayaKrishna2017}. In equilibrium, firms cross-check the messages against each other, and under the asymmetric-correlation information structure, concurrent communication ensures that the signal is the most informative.} In cases where fewer than two legacy carriers serve a market, $\texttt{Capacity-Discipline}_{m,t}$ is set equal to 0. 
While $\texttt{Carrier-Capacity-Discipline}_{j,t}$ varies by carrier and year-month, our treatment $\texttt{Capacity-Discipline}_{m,t}$ varies by market and year-month. This is an important distinction for the empirical analysis, where the observations are at the market-carrier-year-month level.

\begin{figure}[t!] \begin{center} \caption{Prevalence of ``Capacity Discipline'' in Earnings Call Transcripts}
\includegraphics[width=.55\linewidth]{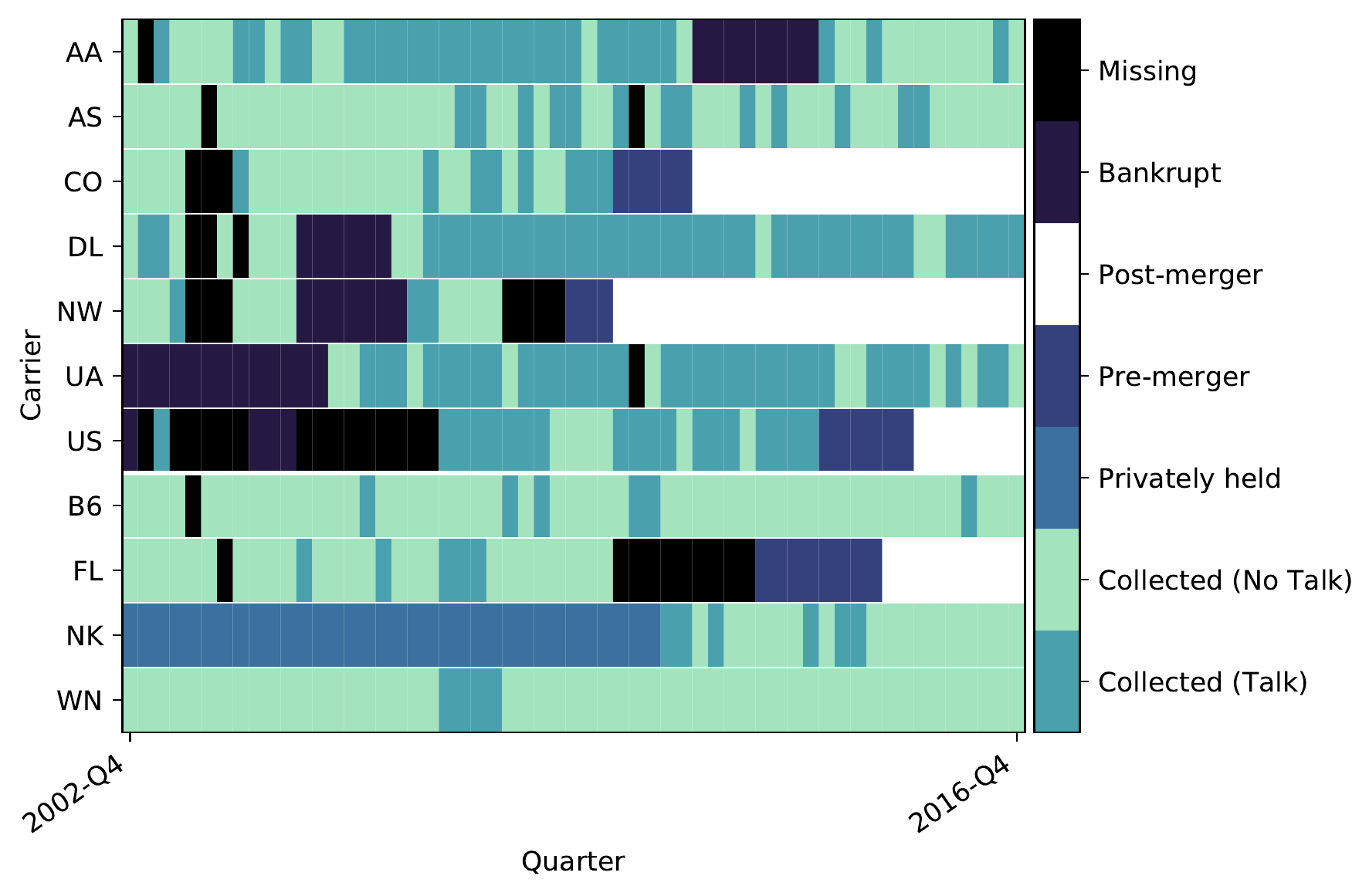}
  \caption*{\footnotesize Notes. This figure shows the availability of transcripts and the prevalence of ``Capacity Discipline'' for 11 airlines. The x-axis denotes years and quarters, and the y-axis denotes the name of the airline. Each color/shade denotes the status of the transcript. Collected (Talk) means the transcript is available and the airline discussed capacity discipline, and Collected (No Talk) means the transcript is available but the airline did not discuss capacity discipline.}
\label{figure:capdis} \end{center} \end{figure}

Figure \ref{figure:capdis} shows the occurrence of $\texttt{Carrier-Capacity-Discipline}_{j,t}$
in our data. Each row corresponds to one airline and shows the periods for which the
carrier discussed capacity discipline.
There is variation in communication across both airlines and time,
which is necessary for the identification.
Even though the reports do not vary within a quarter, the composition of airlines operating in markets---market structure---varies both within a quarter and across quarters, causing the dummy variable $\texttt{Capacity-Discipline}_{m,t}$ to vary by month.

\begin{table}[t!]
\caption{Summary Statistics \label{table:sum-stat}}
\begin{adjustbox}{width=1\textwidth}
	\begin{tabular}{lllllllllllll}\toprule
		& \multicolumn{3}{c}{Seats} & \multicolumn{2}{c}{Cap. Discipline} & \multicolumn{2}{c}{Talk Eligible} & \multicolumn{2}{c}{Monopoly Market} & \multicolumn{2}{c}{Missing Report}&\\
\cmidrule(l{.75em}){2-4} \cmidrule(l{.75em}){5-6}\cmidrule(l{.75em}){7-8}\cmidrule(l{.75em}){9-10}\cmidrule(l{.75em}){11-12}
&Mean&SD&Median&Mean&SD&Mean&SD&Mean&SD&Mean&SD&N \\
\midrule
\textbf{Carrier Type}&&&&&&&&&&&& \\
Legacy&11,757.894&12,264.478&7,364.000&0.089&0.285&0.311&0.463&0.546&0.498&0.262&0.440&562,469 \\
LCC&11,255.056&10,467.260&8,220.000&0.032&0.177&0.106&0.307&0.471&0.499&0.098&0.297&279,522 \\
\midrule
\textbf{Market Participants}&&&&&&&&&&&& \\
Mixed Market&13,349.373&12,749.700&8,990.000&0.058&0.235&0.197&0.398&0.321&0.467&0.147&0.354&410,888 \\
Legacy Market&9,915.007&10,330.230&6,282.000&0.082&0.274&0.287&0.452&0.713&0.452&0.265&0.441&431,103 \\
\midrule
\textbf{Total}&11,590.963&11,700.888&7,776.000&0.070&0.256&0.243&0.429&0.521&0.500&0.207&0.405&841,991 \\
\bottomrule
	\end{tabular}
\end{adjustbox}

\caption*{\footnotesize Notes. Table of summary statistic for all key variables. Observations are at the carrier-market-month level for airport-pair markets.}
\end{table}

\cref{table:sum-stat} provides a summary of this airline data. Legacy carriers offer, on average, 11,757.9 seats in a month, while LCCs offer 11,255.1.\footnote{ We use the seats variable in the T-100 dataset, which corresponds to the scheduled seats transported in a month between two airports. If we use seats weighted by the share of performed departures over scheduled departures, the main empirical findings do not change.} Consistent with our focus on the communication among only the legacy carriers, we find that legacy carriers are far more likely to be in a market where $\texttt{Capacity-Discipline}$ is equal to 1.\footnote{  Despite the lawsuit, we do not include Southwest (WN) when assessing communication because it is known to have a different cost structure and business model than the legacy carriers, and, more importantly, the notion of capacity discipline appears only four times in the entire Southwest's transcripts; see row WN in \cref{figure:capdis}.}

We define the categorical variable $\texttt{Talk-Eligible}_{m,t}\in\{0,1\}$ to be equal to 1 if there are at least two legacy carriers in market $m$ in period $t$ and 0 otherwise. This variable controls for the possibility that markets where legacy carriers \emph{could} engage in coordinating communication may be fundamentally different from markets where such communications are not possible. Not including this control variable would confound the correlation between talking and seats. \cref{table:sum-stat} shows that, on average, 24\% of the observations in our sample have the potential for coordinating communications. In a similar vein, markets served by a single carrier could differ from non-monopoly markets. We account for this possibility by introducing a categorical variable $\texttt{MonopolyMarket}_{m,t}$, which is equal to 1 if in $t$, market $m$ is served by only one firm  and equal to 0 otherwise. We also see that, on average, 52\% of observations are monopoly markets, and that legacy carriers are more likely to serve monopoly markets than LCCs.

As
discussed above, we take special note of markets where we were unable to collect an earnings call
transcript.\footnote{   See \cref{section:ML} for a discussion of when and why we were unable to
collect a transcript. Transcripts are missing, mostly for legacy carriers, largely due to their increased prevalence of bankruptcies and mergers.} To account for that, we introduce a categorical variable $\texttt{MissingReport}_{m,t}\in\{0,1\}$ is equal to 1 if at least one of the legacy carriers serving market $m$ in period $t$ did not hold an earnings call for the quarter prior to month $t$. 
\cref{table:sum-stat} shows that legacy carriers are more likely to operate in a market that is missing a report---a result
of the bankruptcy by many of the legacies. 
Following \citemain{Jones1996}, in our regression, we use \texttt{MissingReport} and its interactions with other covariates to account for missing reports. 

\section{Empirical Analysis}\label{section:empirical}

In this section, we specify and estimate a model to 
 investigate whether the data are consistent with the allegation that U.S. legacy carriers used their quarterly earnings calls to coordinate capacity reductions. 
 We begin with the premise that airlines have access to communication technology (the quarterly earnings call) and posit that such technology allows them to signal to others their intention to coordinate future capacities. In particular, we hypothesize that when all legacy airlines serving a market communicate concurrently (by announcing they will adhere to capacity discipline), it signals to everyone else their intention to reduce capacity, maintaining coordination. For our hypothesis to work, it is essential that every legacy airline in a market simultaneously communicates.

\subsection{Primary Model and Results}\label{sec:primary_results}
We examine the relationship between communication among legacy airlines and the seats they offer between 2003:Q1 and 2016:Q3 (inclusive). We use panel data model to estimate these relationships by estimating the following model using the within-group estimator:

\begin{equation}\label{eq:model_capdis}
	\begin{aligned}
		\ln(\texttt{seats}_{j,m,t})=& \beta_{0} \times \texttt{Capacity-Discipline}_{m,t}+\beta_{1} \times \texttt{Talk-Eligible}_{m,t}\\
		&\qquad+ \beta_{2}\times \texttt{Monopoly}_{m,t}+\beta_{3} \times \texttt{MissingReport}_{m,t}\\
		&\qquad+ \beta_{4}\times \texttt{Talk-Eligible}_{m,t} \times \texttt{MissingReport}_{m,t}\\
		&\qquad+ \beta_{5}\times \texttt{Monopoly}_{m,t} \times \texttt{MissingReport}_{m,t}\\
		&\qquad+ \mu_{j,m} + \mu_{j,yr,q} + \gamma_{origin,t} + \gamma_{destination,t} + \varepsilon_{j,m,t},
	\end{aligned}
\end{equation}
 where the dependent variable is the log of total seats made available by airline $j$ in (airport-pair) market $m$ in month $t$. Our variable of interest is $\texttt{Capacity-Discipline}_{m,t}$, which is the dummy variable introduced in \cref{section:ML} is equal to 1 if there are at least two legacy carriers in market $m$ and month $t$, and they all communicated about capacity discipline in their previous quarter's earnings calls, and 0 otherwise.

The idea behind capacity discipline is that airlines restricted seats even when there was adequate demand, which can vary across both markets and time. 
To control for these unseen factors, we include carrier-market fixed effects, $\mu_{j,m}$, and carrier-year-quarter fixed effects, $\mu_{j,yr,q}$. These fixed effects allow airlines to provide different levels of capacity across different markets and time.  During our sample period, we observe several mergers (see \cref{figure:capdis}). Since it is possible that a carrier's relationship to a specific market could change in a meaningful way after a merger, we redefine the carrier as the merged entity in order to allow greater flexibility in these fixed effects. For example, the fixed effect for American Airlines serving the ITH-PHL market is allowed to differ before and after American merges with US Airways. Lastly, to control for time-dependent changes in demand we use origin- and destination-airport specific time trends, $\gamma_{origin,t}$ and $\gamma_{destination,t}$.\footnote{ Implicitly, we are assuming that our panel data model satisfies the strict-exogeneity assumption. We performed a diagnostic test proposed by \citepmain[page 285]{Wooldridge2001} by including the lead $\texttt{Capacity-Discipline}_{m,t+1}$ as an additional regressor. This regressor's estimated coefficient was +0.007 and statistically significant at the 5 percent level, which suggests that the assumption of strict exogeneity is reasonable in our context.}

Next, we explain the identification strategy for \cref{eq:model_capdis}. To highlight the key sources of variation in the data, we fix an airline---say, Delta (i.e., $j=DL$)---and consider different potential market structures and communication scenarios in Table \ref{table:identification}. In markets $m=1,2$, only DL operates, so the concept of communication is moot and $\texttt{Capacity-Discipline}_{1,t}=\texttt{Capacity-Discipline}_{2,t}=0 $. Then we can use variation in whether a report is available (for $m=2$) or not (for $m=1$) to identify $\beta_{2}$ and $\beta_{3}$, as shown in the last column. Market $m=3$ is served by both DL and UA and both discuss ``capacity discipline'' in the previous quarter, so $\texttt{Capacity-Discipline}_{3,t}=1$, which identifies $\beta_{0}+\beta_{1}$. The same identification argument applies to identifying $\beta_{0}+\beta_{1}$ in markets $m=6,7$ where every airline in the market talks and a report for DL is available, even when an LCC is present ($m=7$). In contrast, for market $m=4$, even when both US and UA discuss capacity discipline, we identify $\beta_{1}+\beta_{3}$ because DL did not have a transcript.

\begin{table}[t!] \centering \caption{Identification of the Parameters\label{table:identification}}
\resizebox{\textwidth}{!}{ \begin{tabular}{lllllllll} \toprule {\bf market} & {\bf market structure} &
{\bf DL reports} & {\bf communicating}            & \texttt{Cap-Dis} & \texttt{Report} &
\texttt{Monopoly} & \texttt{Talk-Eligible} & {\bf parameters} \\ \midrule 1   & \{DL\} & no & n/a &
0    & 1   & 1 & 0 & $\beta_3+\beta_2$        \\ 2 & \{DL\} & yes    & n/a & 0 & 0
& 1    & 0       & $\beta_2$ \\ 3 & \{DL, UA\}     & yes    & \{DL, UA\} & 1 & 0
 & 0    & 1       & $\beta_{0}+\beta_{1}$ \\ 4   & \{DL, UA, US\}   & no    &
\{US\} or \{UA\} or \{US, UA\} & 0    & 1   & 0    & 1 & $\beta_{3}+\beta_{1}$
\\ 5 & \{DL, UA, US\}   & yes & \{US, UA\}         & 0 & 0   & 0    & 1 &
$\beta_{1}$ \\ 6 & \{DL, UA, US\} & yes    & \{DL, UA, US\} & 1    & 0 & 0 & 1       &
$\beta_{0}+\beta_{1}$      \\ 7   & \{DL, UA, US, F9\} & yes & \{DL, UA, US\} & 1    &
0   & 0    & 1       & $\beta_{0}+\beta_{1}$ \\ 8   & \{DL, F9\} & yes    & n/a &
0 & 0   & 0    & 0 & - \\ \bottomrule \end{tabular}} \caption*{\footnotesize Notes. An example to
show identification from the perspective of Delta, i.e., when $j=DL$, and here UA and US are legacy
carriers while F9 is an LCC.
} \end{table}

Lastly, we identify the fixed effects using the deviation from the mean.
Therefore, the key source of identification is the variation in \texttt{Capacity-Discipline} across markets and over time (see 
Figure \ref{figure:capdis}), which in turn depends on the variation in market structure and communication. 
We also assume that conditional on all control variables, \texttt{Capacity-Discipline} is uncorrelated with the error, and this conditional exogeneity of treatment is sufficient to identify the relationship between \texttt{Capacity-Discipline} and log-seats \citepmain{Rosenbaum1984}.

 We present the estimation of the semi-elasticity from \cref{eq:model_capdis} in column (1) of \cref{tab.main_results}.\footnote{ Throughout the paper, for a binary regressor, we present its estimated semi-elasticity. If the estimated coefficient of a dummy variable in a semilogarithmic regression is $\hat{\beta}$, then the effect of the dummy variable on the outcome variable is $100 \times (\exp(\hat{\beta}) - 1)\%$ \citepmain{HalvorsenPalmquist1980}.\label{footnote:correction}} Using our model, we find that when all of the legacy carriers in a talk-eligible market communicate with each other about capacity discipline, there is a subsequent reduction in the number of seats offered by an average of 2.02\%.\footnote{This estimate is a weighted average of the parameter estimates across markets, time, and types of carriers and should be interpreted as a percentage decrease in capacities. Following \citemain{ChaisemartinDHaultfuille2020}, we estimated the weights for each group, and only 0.07\% of those weights were negative, suggesting that negative weights do not drive our estimate.} The standard errors are clustered at the bi-directional market level.

%\begin{landscape}
\begin{table}[t!] \centering \caption{Communication and Available Seats}\label{tab.main_results} \begin{threeparttable} \scalebox{0.7}{
{
\def\sym#1{\ifmmode^{#1}\else\(^{#1}\)\fi}
\begin{tabular}{l*{6}{c}}
\toprule
                    &\multicolumn{1}{c}{(1)}&\multicolumn{1}{c}{(2)}&\multicolumn{1}{c}{(3)}&\multicolumn{1}{c}{(4)}&\multicolumn{1}{c}{(5)}&\multicolumn{1}{c}{(6)}\\
                    &\multicolumn{1}{c}{Log Seats}&\multicolumn{1}{c}{Log Seats}&\multicolumn{1}{c}{Log Seats}&\multicolumn{1}{c}{Log Seats}&\multicolumn{1}{c}{Log Seats}&\multicolumn{1}{c}{Log Seats}\\
\midrule
Capacity Discipline &     -0.0202&     -0.0150&            &            &            &            \\
                    &    (0.0060)&    (0.0049)&            &            &            &            \\
Capacity Discipline 2&            &            &     -0.0193&     -0.0160&            &            \\
                    &            &            &    (0.0064)&    (0.0052)&            &            \\
Capacity Discipline 3&            &            &     -0.0285&     -0.0109&            &            \\
                    &            &            &    (0.0112)&    (0.0103)&            &            \\
Capacity Discipline 4&            &            &     -0.0332&     -0.0018&            &            \\
                    &            &            &    (0.0541)&    (0.0342)&            &            \\
Legacy Market x Capacity Discipline&            &            &            &            &     -0.0169&     -0.0197\\
                    &            &            &            &            &    (0.0074)&    (0.0063)\\
Mixed Market x Capacity Discipline (Legacy)&            &            &            &            &     -0.0195&      0.0010\\
                    &            &            &            &            &    (0.0115)&    (0.0085)\\
Mixed Market x Capacity Discipline (LCC)&            &            &            &            &     -0.0341&     -0.0221\\
                    &            &            &            &            &    (0.0155)&    (0.0107)\\
\midrule
Carrier-Market FE's &$\checkmark$&            &$\checkmark$&            &$\checkmark$&            \\
Carrier-Market-Structure FE's&            &$\checkmark$&            &$\checkmark$&            &$\checkmark$\\
R-squared           &       0.088&       0.083&       0.088&       0.083&       0.088&       0.084\\
N                   &     841,991&     841,991&     841,991&     841,991&     841,991&     841,991\\
\bottomrule
\end{tabular}
}

}\label{table:main}
\caption*{\footnotesize Notes. We report semi-elasticities (see \cref{footnote:correction}), with standard errors clustered at the bi-directional market level in parentheses. Other control variables included in all regressions, but whose coefficients are not reported are \texttt{Talk-Eligible}, \texttt{Monopoly},  \texttt{MissingReport}, interactions of the \texttt{MissingReport} indicator with \texttt{Talk-Eligible} and \texttt{Monopoly}. In columns 2 and 3, these coefficients are allowed to vary based on the number of legacy carriers in the market (either 0 or 1, 2, 3, 4, or 5 legacy carriers). In columns 5 and 6, these coefficients are allowed to vary across legacy and mixed markets, and within mixed markets for legacy carriers and LCCs. Additionally, all regressions include origin- and destination-airport annual time trends, and carrier-year-quarter fixed effects. Columns 1, 3, and 5 include carrier-market fixed effects, and columns 2, 4, and 6 include carrier-market-structure fixed effects.
}
\end{threeparttable} \end{table}

To determine the estimate's economic significance, we can compare it to the average change in capacities in a set of relevant comparison markets. To do so, we identify all talk-eligible markets where communication did not occur. In other words, we define our set of comparison markets as those where communication \emph{could} have occurred, but did not. In such markets, we find that the average percentage change in capacities is 3.67\%. So, whenever legacy airlines communicate, their capacities drop by more than 50\% of the average change in capacities in our comparison markets, a significant reduction.

While we attempt to capture some of the differences in market structures that permit communication (via the \texttt{Talk-Eligible} variable), this may not adequately capture the manner in which competitive behavior may respond to market structure, either in terms of the number or type of carriers, or the specific set of carriers serving a market.\footnote{In our primary specification, identification of $\beta_0$ relies on variation in communication and/or market structure, as \texttt{Capacity-Discipline} can turn on or off as a result of carriers beginning or ending communication, or when a legacy carrier enters or leaves a market. In our data, we find that 85.4\% of changes in \texttt{Capacity-Discipline} derive exclusively from changes in communication, while 14.6\% of changes occur concurrently with changes in market structure.}  To address this concern, we re-estimate our primary specification \cref{eq:model_capdis}, but control for specific market structures. In particular, we change the carrier-market fixed effects in \cref{eq:model_capdis} to carrier-market-\emph{structure} fixed effects.

To best understand the carrier-market-structure fixed effects, consider an example of the Ithaca (ITH) to Philadelphia (PHL) market. Suppose we observe this market for four periods, and during this time the market structures are \{AA, DL\}, \{AA\}, \{AA, UA\}, and \{AA, DL\}. The carrier-market fixed effects for a given carrier would be constant across all periods in which they compete in the ITH-PHL market, but carrier-market-structure fixed effects allow American (AA) to behave differently when in a duopoly with Delta (DL) compared to when it is competing in a duopoly with United (UA). In \cref{tab.main_results}-column (2) we present the estimation results from this alternative specification. We find that communication is associated with a 1.50\% reduction in offered capacity.\footnote{ Under the carrier-market-structure fixed effects, \texttt{Talk-Eligible} and \texttt{Monopoly} are redundant and are therefore excluded from the regressions.}

Next, we consider whether the relationship between communication and capacity varies with the number of communicating airlines. 
Let $\texttt{Capacity}\texttt{-Discipline-}k_{m,t}\in\{0,1\}$ be 1 if market $m$ in period $t$ is talk eligible, is served by exactly $k$ legacy carriers, and all $k$ of them use capacity discipline.
Then we estimate \cref{eq:model_capdis} after replacing $\texttt{Capacity-Discipline}_{m,t}$ with three (additively separable) indicators $\{\texttt{Capacity}\texttt{-Discipline-}k_{m,t}: k=2,3,4\}$. 
The estimation results using the carrier-market fixed effects and the carrier-market-structure fixed effects are in columns (3) and (4) of \cref{tab.main_results}, respectively.\footnote{ While we do observe some markets with five legacy carriers, $\texttt{Capacity-Discipline}_{m,t}$ is always zero in these markets, and so we do not include an additional variable for this case.}

In column (3) of \cref{tab.main_results}, we find that with the carrier-market fixed effects, the association between communication and capacity reductions are increasing in the number of legacy carriers serving the market. In particular, we find that communication is associated with a reduction in capacity of $1.93\%$, $2.85\%$ and $3.32\%$ in markets with two, three and four legacy carriers, respectively. Although, because there are few markets with four legacy carriers, this coefficient is imprecisely estimated. 
With the carrier-market-structure fixed effects, however, we find that communication is associated with a reduction in capacity by $1.60\%$ in markets with two legacy carriers. For the markets with three or four legacy carriers, the coefficients are imprecisely estimated with no effect.

Lastly, we explore how the estimate change between markets with only legacy carriers and mixed markets with both legacy and LCCs. We present summary statistics for these two types of markets in \cref{table:sum-stat}. 
We present the results from this exercise using the carrier-market fixed effects and the carrier-market-structure fixed effects in columns (5) and (6) of \cref{tab.main_results}, respectively. 
With the carrier-market fixed effects, we find that communication about capacity discipline is associated with a 1.69\% decrease in the number of seats offered. 
In mixed-markets, we find that communication is associated with a 1.95\% decrease in legacy seats and a 3.41\% decrease in LCCs seats. 

In summary, we find that capacity is lower when all legacy carriers serving a talk-eligible market discuss capacity discipline, a finding which is consistent with the allegation that U.S. legacy carriers used their quarterly earnings calls to coordinate capacity reductions. On average, we find that capacity is between 1.50\% and 2.02\% lower when this communication occurs, though we find this varies with the number of legacy carriers in a market, and the presence of LCCs.\footnote{ In Appendix \ref{sec:market_hetero} we explore how the relationship between capacity and communication varies with the size of a market and the amount of business travel in a market.}

In the analysis that follows, we use the specification outlined in \cref{eq:model_capdis}, which employs carrier-market fixed-effects, as our primary specification because it takes advantage of both important sources of variation in \texttt{Capacity-Discipline}, namely, that \texttt{Capacity-Discipline} can turn on or off as a result of a legacy carrier beginning or ending communication, or when a legacy carrier enters or leaves a market.\footnote{ An additional concern with including carrier-market-structure fixed effects is that market structure may correlate with the unobservable in \cref{eq:model_capdis}. As noted in \cref{section:robustness}, and with more details in Appendix \ref{sec:cf}, using a control function approach we find that our primary results are robust to this concern.} 
	For completeness, we provide corresponding estimates in Appendix \ref{section:additional_fe} for everything that follows using carrier-market-structure fixed effects.

\subsection{Market-Level Changes in Capacity and Number of Flights}
 To complement our previous analysis, we examine the relationship between capacity and communication among legacy airlines at the market level. In particular, we ask whether the firm-level reductions in seats that we have estimated above involve a reduction in total market capacity, a reduction in the number of scheduled flights, or both.

To shed light on the first question, we aggregate capacity to the market level, and estimate the same panel data model as \cref{eq:model_capdis}. In this case, the dependent variable is the sum of all seats offered by all the carriers in market $m$ in period $t$. We present the estimation of the semi-elasticity from this model in column (1) of \cref{tab:congestion_results}. We find that overall capacity is lower by 1.7\% when communication about capacity discipline occurs, suggesting that reductions in individual-level capacities translates into a market-wide reduction in offered seats. 

Next, we estimate the relationship between communication and the number of flights in a market. 
To this end, we assume that the number of flights in a market is a Poisson random variable, where the mean depends on all the explanatory regressors in \cref{eq:model_capdis}, including the fixed effects. 
Let $Y_{mt}$ and ${\bf X}_{mt}$, respectively denote the number of flights and observed characteristics (i.e., the right-hand side terms in \cref{eq:model_capdis}) in market $m$ in period $t$; and let $\gamma_m$ and $\gamma_t$ be the market-$m$ fixed effects and time-$t$ fixed effects, respectively. We assume that the probability $Y_{mt}=y$, given $(\gamma_m, \gamma_t)$ and ${\bf X}_{mt}$,  is given by  
\begin{eqnarray}
\Pr(Y_{mt}=y| \gamma_{m},\gamma_{t}, {\bf X}_{mt}) = \frac{\exp(-\mu(\gamma_{m},\gamma_{t}, {\bf X}_{mt}))\left[\mu(\gamma_{m}, \gamma_{t},{\bf X}_{mt})\right]^{y}}{y!},\label{eq:poisson_flight}
\end{eqnarray}
with the mean function $\mathbb{E}(Y_{mt}|\gamma_{m},\gamma_{t}, {\bf X}_{mt})=\mu(\gamma_{m},\gamma_{t}, {\bf X}_{mt})=\exp(\gamma_m +\gamma_t+ {\bf X}_{mt}{\bf \beta})$. 
We can then use the method of conditional maximum likelihood to estimate $\beta$; see, for example, \citemain{Wooldridge1999} and \citemain{Wooldridge2001}, Chapter 18.2.

The estimate of the coefficient for capacity discipline from Eq. (\ref{eq:poisson_flight}) is in column (2) of \cref{tab:congestion_results}. Here too, we find evidence that discussion of capacity discipline is associated 0.02 fewer flights, on average. Thus, we find evidence that, on average, there are fewer total seats offered and fewer flights flown in markets when carriers discuss capacity discipline.
 
\begin{table}[t!]\centering
\caption{Communication, Market-Level Capacity, Flights, Departure Crowding, and Prices}\label{tab:congestion_results}
\begin{threeparttable} 

\scalebox{0.78}{
{
\def\sym#1{\ifmmode^{#1}\else\(^{#1}\)\fi}
\begin{tabular}{l*{5}{c}}
\toprule
                    &\multicolumn{1}{c}{(1)}&\multicolumn{1}{c}{(2)}&\multicolumn{1}{c}{(3)}&\multicolumn{1}{c}{(4)}&\multicolumn{1}{c}{(5)}\\
                    &\multicolumn{1}{c}{Mkt. Capacity}&\multicolumn{1}{c}{Flights}&\multicolumn{1}{c}{Depart. Crowding}&\multicolumn{1}{c}{Price}&\multicolumn{1}{c}{Price}\\
\midrule
Capacity Discipline &     -0.0170&     -0.0221&      0.0501&      0.0061&            \\
                    &    (0.0087)&    (0.0053)&    (0.0290)&    (0.0033)&            \\
Capacity Discipline x Log Market Seats&            &            &     -0.0043&            &            \\
                    &            &            &    (0.0026)&            &            \\
Capacity Discipline (Legacy)&            &            &            &            &      0.0049\\
                    &            &            &            &            &    (0.0035)\\
Capacity Discipline (LCC)&            &            &            &            &      0.0186\\
                    &            &            &            &            &    (0.0057)\\
Log Market Seats    &            &            &      0.0743&            &            \\
                    &            &            &    (0.0031)&            &            \\
\midrule
R-squared           &       0.172&           .&       0.077&       0.133&       0.138\\
Prob $>$ chi2       &           .&       0.000&           .&           .&           .\\
N                   &     614,311&     614,256&     463,951&     649,204&     649,204\\
\bottomrule
\end{tabular}
}

}
\caption*{\footnotesize Notes. The table displays estimated coefficients from the market-level analysis. Column (1) shows the estimate coefficient for the total number of seats offered in each market; column (2) shows the estimate coefficient from the Poisson model on the number of scheduled flights; column (3) shows the estimate coefficient for Departure Crowding, which refers to the average difference between two flights' departure times within an airport; and columns (4) and (5) show the estimate coefficient from price regression where prices are the log of average fares. For all columns, except (2), we report semi-elasticities (see \cref{footnote:correction}) and standard errors clustered at the bi-directional market level in parentheses. For column (2), we report the coefficient $\hat{\beta}$ and robust standard error. Other control variables included in all regressions, but whose coefficients are not reported are are \texttt{Talk-Eligible}, \texttt{Monopoly}, \texttt{MissingReport}, interactions of the \texttt{MissingReport} indicator with \texttt{Talk-Eligible} and \texttt{Monopoly}, origin- and destination- airport annual time trends, and carrier-market fixed effects. Columns (1), (2), and (3) include year-quarter fixed effects, and columns (4) and (5) include carrier-year-quarter fixed effects.}
\end{threeparttable}
\end{table}

\subsection{Crowding of Flight Departure Times and Prices}\label{sec:welfare}
We now turn our attention to other, consumer-welfare relevant outcomes, associated with (i) the preferences of consumers for the time of departure; and, (ii) with the possibility of higher prices.

First, reductions in capacity (relative to demand) could allow airlines to better coordinate the timing of flights. The consumer welfare impact of such coordination is ex-ante ambiguous. On one hand, this could benefit consumers who value greater product differentiation by providing flights at times closer to their preferred time of departure. Additionally, it could reduce the level of congestion (and of associated delays) at capacity constrained airports. On the other hand, such coordination might negatively impact consumers if the distribution of consumers' preferences for travel times is concentrated around a small number of times in the day (e.g., 7am and 6pm for daily business travelers), or if consumers have preferences for short layovers when making a connecting flight.

Second, the capacity reductions might not ultimately affect prices, thus limiting the impact on consumer welfare.\footnote{  For instance, \citemain{ArmantierRichard2003} consider the effect of information exchanges between UA and AA out of O'Hare airport and find that while airlines benefit, it only moderately hurts consumers. They conclude, ``Hence, a marketing alliance between AA and UA, with the sole objective of exchanging cost information, would be advantageous to airlines without significantly hurting consumers."}

Estimating the welfare effect of communication is well beyond the scope of this paper, but we can determine (i) if conditional on reducing capacity, airlines change their departure times and, thereby reduce the crowding of flight departure times; and (ii) if communication is associated with higher average fares.

As we show next, we find no evidence to support the hypothesis that departure crowding has changed, and we find evidence that fares may have risen, both of which show that capacity discipline likely hurt consumers. We show our results next.

\subsubsection{Crowding of Flights Departure Times}
First, we examine if, conditional on reducing capacity, legacy airlines change their departure times and reduce the extent to which flights are scheduled at the same time at the airport.
 To measure the crowding of flight departures times at an airport, we use the following measure proposed by \citemain{BorensteinNetz1999}.

On a route with $n$ daily departures departing $d_1, \ldots, d_n$ minutes after the midnight, the average time difference between two flights is given by 
\begin{eqnarray*}
\texttt{Average-Time-Difference}:=\frac{2}{n-1}\sum_{i=1}^n\sum_{j>i}^n\sqrt{\min\{|d_i-d_j|, 1440-|d_i-d_j|\}}.
\end{eqnarray*}
To make this measure comparable across markets with different $n$, we normalize it by the maximum time difference if the flights were equally spaced throughout a day, such that values close to 1 corresponds to the least crowded flights. Although we use the normalized measure, for notational ease, we continue to refer it as \texttt{Average-Time-Difference}. To calculate \texttt{Average-Time-Difference} we use the Bureau of Transportation Statistics's Airline On-Time Performance database, which records flight times.
  
We estimate a fixed effects model, where the dependent variable is \texttt{Average-Time-Difference} and the regressors are the same as in \cref{eq:model_capdis}, plus two additional variables: the total log-seats offered in the market and an interaction term between the total log-seats and \texttt{Capacity-Discipline}. 
Departure crowding is at the market-level, so we replace the carrier-market fixed effect with market fixed effects.
Our primary variable of interest is the interaction term because it estimates the relationship between log-seats and the changes in the average time difference with communication. If, conditional on reducing offered seats, airlines were increasing the average time between their flights and reducing crowding, then this interaction term's coefficient would be positive. 

We present the estimation results in column (3) of \cref{tab:congestion_results}, under the heading ``Depart. Crowding.'' As we can see, the coefficient for the interaction term is $-0.0043$, but is imprecisely estimated, suggesting that there is no evidence to support the claim that conditional on reducing offered seats, communication is associated with less crowded departures. 

\subsubsection{Ticket Prices}\label{section:prices}
Next, we consider estimating the relationship between communication and prices. If, whenever airlines communicate, they lower their offered capacities, then, unless capacities never bind, it is reasonable to expect that prices would rise due to communication.

Even though it might seem straightforward to estimate this relationship, for example, by estimating \cref{eq:model_capdis} after replacing the log of offered seats as the dependent variable with the log of the prices as the dependent variable, this empirical strategy is infeasible.  Airlines sell tickets for origin to \emph{final}-destination pairs, but the offered capacities and communication are at the direct-segment level. Thus, to understand the relationship between communication and prices, we must first construct a new dataset of prices and communication.

Connecting tickets involve flights that go through different nonstop segments, possibly with different market structures in each segment.
Thus, while the prices are at the origin-destination level, capacity plans and our communication measure are at the nonstop segments level. So,  we have to aggregate capacity and communication from the segment level to the origin-destination level. 
For example, consider flights traveling from A to C via a connecting airport, B. In particular, assume that in segment A--B, two airlines are talking, but in segment B--C, there are three airlines but the third airline is not talking. Our aggregation must account for how to define \texttt{Capacity-Discipline} in these and similar situations. 
Furthermore, airlines may use multiple routes for the same market (i.e., use multiple airports to connect a given origin and destination), adding additional complexity to our problem. 

Next, we define how we aggregate communication in segments A--B and B--C to determine communication in the origin-destination pair A to C.
First, we follow \citemain{Borenstein1989} and construct a dataset of prices, but instead of aggregating at the market level (e.g., market A to C), we aggregate them at the market-route level. For example, consider a ticketing carrier, say UA, serving A to C via two routes, AB-BC and AB-BD-DC. In this case, we treat these two routes separately, even though they are have the same origin and destination. At the end of this aggregation, we have average prices and the total number of passengers transported by each airline for each market-route. We then use the number of passengers transported to determine weighted average prices and \texttt{Capacity-Discipline}, weighted by the number of passengers in those combinations defined at the carrier-market level.

In particular, to determine $\texttt{Capacity-Discipline}_{m,t}$ at the route level, we calculate \texttt{Capacity-Discipline} for every nonstop segment. Then we merge the price data with these new communication data and restrict the sample in the price data to those markets we observe in our primary analysis.\footnote{  For instance, as ITH-CHO is not served nonstop by any airline, it does not appear in our primary analysis. We drop this market from this analysis, even though there are connecting flights between them.} Note that the number of carriers serving a market in our price dataset weakly exceeds the number of carriers serving that market in our primary analysis because they include carriers that serve the origin and destination pair via a connection.

 We can then aggregate the dummy variable \texttt{Capacity-Discipline} that we defined previously from the segment level to the origin-destination level. 
 In particular, if the variable $\texttt{Capacity-Discipline}=1$ in all nonstop segments of a route, then we define $\texttt{Capacity-Discipline}=1$ for that route. 
 For the market missing report variable, we take the opposite approach: if it is 1 for at least one segment, then it is 1 for the route.
Finally, we construct a \texttt{Capacity-Discipline} variable for each market by taking the passenger weighted average of \texttt{Capacity-Discipline} for each route through which a carrier serves that market.

To better understand this approach, consider the following stylized example. Suppose a carrier serves a market-quarter $\{m,t\}$ via three different routes, and \texttt{Capacity-Discipline} variable is $1, 0,$ and $1$ for these three routes. Furthermore, if the carrier sends 25\% of its passengers along route 1, 25\% along route 2, and 50\% along route 3, then \texttt{Capacity-Discipline} variable for the carrier in $\{m,t\}$ is equal to $1\times 0.25 + 0 \times 0.25 + 1\times 0.5=0.75$. We use the same approach to calculate the \texttt{Talk-Eligible}, \texttt{Monopoly},  and \texttt{Missing-Report} variables.

Using these variables in a panel data model like \cref{eq:model_capdis} we estimate the relationship between \texttt{Capacity-Discipline} and the log of (average) route-level prices. The results are in columns (4) and (5) in \cref{tab:congestion_results}. 
The estimates suggest that the average price increased by 0.59\%, and that this increment is mostly due to LCCs, whose average prices increased by 1.80\%. 

In summary, we find no evidence that the crowding of flight departure times changed, and we find evidence that prices may have risen.\footnote{ Our analysis treats capacity choices as strategic substitutes. It is reasonable to consider the possibility that if consumers care about departures, and if this preference is strong enough, that may soften competition to the effect that the capacity choices become strategic complements and not strategic substitutes. However, we do not believe this to be the case because airlines' departures and capacity choices are interlinked. Thus, even after setting aside airlines' communication decisions, we would have to consider three choices (departure times, capacity choices, and airfares) together in our model. There are several ways to model departures. One of them is the Salop/Vickrey circular city model, as in \citemain{GuptaLaiPalSarkarYu2004}, which is also consistent with \citemain{BorensteinNetz1999}. We can then embed this model within a \citemain{KrepsScheinkman1983} framework, which results in a game played by airlines in three stages. First, they choose departure times, then the capacities and the prices. However, conditional on the circle locations (i.e., the departure times), capacities are still strategic substitutes. \label{footnote:KrepsScheinkman}}

\section{Robustness Exercises}\label{sec:robustness}

In \cref{section:empirical}, we found that whenever all of the legacy carriers in a market discuss capacity discipline, capacity is on average 2\% lower in the next quarter, a finding which is consistent with the accusation that legacy carriers used their earnings calls to coordinate with other carriers to reduce capacity. In this section, we perform a series of robustness exercises to address other possible explanations for this finding. 

\subsection{Financial Transparency or Coordination}\label{sec:financial_transparency}

We have shown that we observe lower capacity when all legacy carriers in a market discuss capacity discipline. Of course, it could be that airlines are not coordinating but are simply announcing their unilateral intentions to reduce capacity in response to demand forecasts or for other reasons specific to themselves. That is, the airlines may be using the quarterly earnings call for its ostensible purpose: to inform investors about the state of their businesses.

If this is the case, then the number of seats offered by an airline would also fall when the airline is communicating, but its competitors are not. That is not what we find. We do not find evidence that a carrier reduces capacity when it discusses capacity discipline, but its legacy competitors do not. Additionally, carriers do not reduce capacity in monopoly markets, where we would also expect to find capacity reductions following communication. Finally, we find no evidence of capacity reductions when all but one of the legacy carriers serving a market discuss capacity discipline.

To investigate whether airlines decrease capacity when they are the only one discussing capacity discipline, we estimate the following variation of \cref{eq:model_capdis}:
\begin{equation} \label{eq:model_only_j}
	\begin{aligned}
		\ln(\texttt{seats}_{j,m,t})=& \beta_{0} \times \texttt{Only-j-Talks}_{j,m,t} + \beta_{1} \times \texttt{Talk-Eligible}_{m,t}\\
		&+ \beta_{2}\times \texttt{Monopoly}_{m,t}+\beta_{3} \times
		\texttt{MissingReport}_{m,t} \\& +{\boldsymbol \beta}_4^{\top}\times \texttt{MissingReport-Interactions}_{m,t}\\&+ \mu_{j,m} + \mu_{j,yr,q} + \gamma_{origin,t} + \gamma_{destination,t} + \varepsilon_{j,m,t},
	\end{aligned}
\end{equation}
where our variable of interest is $\texttt{Only-j-Talks}_{j,m,t}$ defined as 
{\small\begin{equation*}
\begin{aligned}
\texttt{Only-j-}&\texttt{talks}_{j, m,t}=\\
&\begin{cases}
	\mathbbm{1}\left\{\texttt{Carrier-Capacity-Discipline}_{j,t}=1 \right. &\\
	 \left.\qquad \wedge\; \texttt{Carrier-Capacity-Discipline}_{k,t}=0\right.&|J_{m,t}^{\texttt{Legacy}}| \ge 2\\
	 \left.\qquad \forall k \ne j \in J_{m,t}^{\texttt{Legacy}}\right\}& \\
	0& |J_{m,t}^{\texttt{Legacy}}| < 2.
\end{cases}
\end{aligned}
\end{equation*}}
\noindent That is, $\texttt{Only-j-Talks}_{j,m,t}$ indicates whether carrier $j$ is the only legacy carrier in market $m$ that discussed capacity discipline, conditional on there being at least two legacy carriers. The parameter $\beta_{1}$ shows the extent to which a legacy carrier that discusses capacity discipline when none of its market-level competitors discussed capacity discipline changes capacity. If discussion of capacity discipline is meant to inform investors about future strategic behavior, $\beta_{1}$ should be negative and, likely, close to -2.02\%. 
We present the estimation results from \cref{eq:model_only_j} in column (1) of \cref{table:robustness}. As we can see from the estimates in the first row of column (1), there is no evidence of a decline in the capacity associated with the unilateral discussion of capacity discipline. We find the opposite: the number of offered seats is 1.72\% higher when airlines communicate unilaterally.

\begin{table}[t!] \centering \caption{Financial Transparency and Information Sharing}\label{table:robustness}
\begin{threeparttable} \scalebox{0.9}{
{
\def\sym#1{\ifmmode^{#1}\else\(^{#1}\)\fi}
\begin{tabular}{l*{5}{c}}
\toprule
                    &\multicolumn{1}{c}{(1)}&\multicolumn{1}{c}{(2)}&\multicolumn{1}{c}{(3)}&\multicolumn{1}{c}{(4)}&\multicolumn{1}{c}{(5)}\\
                    &\multicolumn{1}{c}{Log Seats}&\multicolumn{1}{c}{Log Seats}&\multicolumn{1}{c}{Log Seats}&\multicolumn{1}{c}{Log Seats}&\multicolumn{1}{c}{Log Seats}\\
\midrule
Only $j$ Talks      &      0.0172&            &            &            &            \\
                    &    (0.0066)&            &            &            &            \\
Monopoly Capacity Discipline&            &      0.0260&      0.0027&            &            \\
                    &            &    (0.0074)&    (0.0050)&            &            \\
Capacity Discipline $N-1$&            &            &            &     -0.0022&            \\
                    &            &            &            &    (0.0045)&            \\
Capacity Discipline ``Not $j$''&            &            &            &            &     -0.0010\\
                    &            &            &            &            &    (0.0074)\\
\midrule
R-squared           &       0.088&       0.088&       0.061&       0.087&       0.087\\
N                   &     841,991&     841,991&     438,980&     841,991&     841,991\\
\bottomrule
\end{tabular}
}

}
\caption*{\footnotesize Notes. Notes. We report semi-elasticities (see \cref{footnote:correction}), with standard errors clustered at the bi-directional market level in parentheses. Other control variables included in all regressions, but whose coefficients are not reported are \texttt{Talk-Eligible}, \texttt{Monopoly},  \texttt{MissingReport}, interactions of the \texttt{MissingReport} indicator with \texttt{Talk-Eligible} and \texttt{Monopoly}, origin- and destination-airport annual time trends, carrier-year-quarter fixed effects, and carrier-market fixed effects. Column 3 omits the \texttt{Talk-Eligible} and \texttt{Monopoly} variables.
}
\end{threeparttable}
\end{table}
A second approach to addressing the concern mentioned above is to look at capacity decisions in monopoly markets. If carriers discuss capacity discipline to inform investors about their plans to reduce capacity, presumably independent of what other airlines are doing, we should expect to see reductions in monopoly markets following those discussions. To estimate the role of ``monopoly capacity discipline" we estimate our primary model \cref{eq:model_capdis}, but using the treatment $\texttt{Monopoly-Capacity-Discipline}_{m,t}$, which is equal to 1 when a carrier in a monopoly market discussed capacity discipline and $0$ otherwise. We estimate this model using both our full sample and a sample that consists of only monopoly markets, and present the results in columns (2) and (3) of \cref{table:robustness}, respectively. In the full sample we find the opposite---capacities are higher after a monopoly airline discusses capacity discipline---but for the monopoly markets sample we find no evidence of an effect.

Finally, we consider whether carriers reduce capacity in cases where all but one of the legacy carriers serving the market discuss capacity discipline. To do so, we estimate \cref{eq:model_capdis} with the treatment variable $\texttt{Capacity-Discipline-N-1}_{m,t}$ defined as 
\begin{equation}
	\begin{aligned}
		\texttt{Ca}&\texttt{pacity}\texttt{-Discipline-N-1}_{m,t}=\\
		&\begin{cases}
			\sum\limits_{j \in J_{m,t}^{\texttt{Legacy}}}\mathbbm{1}\left\{\texttt{Carrier-Capacity-Discipline}_{j,t}\right\} = |J_{m,t}^{\texttt{Legacy}}| - 1 & ,|J_{m,t}^{\texttt{Legacy}}| \ge 2\\
		0& ,|J_{m,t}^{\texttt{Legacy}}| < 2,\\
		\end{cases}
	\end{aligned}\label{eq:n-1}
\end{equation}
which is equal to 1 when all but one of the legacy carriers in a $\texttt{Talk-Eligible}$ market discuss capacity discipline, and 0 otherwise. 
We present this estimation results in column (4) of \cref{table:robustness}. We find no evidence of a relationship between communication and capacity when all but one of the legacy carriers serving a market discuss capacity discipline. In light of these exercises---looking at markets where one carrier speaks but its competitors do not, looking at capacity decisions in monopoly markets, and looking at markets where all but one legacy carrier speak---we conclude that discussion of capacity discipline is not simply a bona fide announcement of future, unilateral intentions.

\subsection{Information Sharing}
So far, we have shown that when all legacy carriers in a market discuss capacity discipline, capacity is subsequently lower, and, if any one of the legacy carriers is not discussing capacity discipline while the others are, their number of offered seats does not change (\cref{table:robustness}, column (4)). While these two results are consistent with coordination, they could also be consistent with the idea that (for some historical reason) airlines use correlated strategies. That is, when they announce their intention to engage in capacity reduction during the earnings call, they share their private information about the aggregate airline demand. 

In fact, our previous finding that the level of capacity reduction is increasing in the number of legacy carriers serving the market (\cref{table:main}, columns (3) and (4)) provides suggestive support for such an alternative hypothesis: when more airlines are communicating, the precision of the aggregate signal gets better, which in turn induces stronger correlation in capacity choices. Thus, this alternative ``information sharing'' model interprets the communication as being payoff relevant, unlike in \citemain{AwayaKrishna2016} wherein capacity discipline is cheap talk, but it \emph{does not} require firms to coordinate on any action.

To better understand this alternative theory, consider the following. Suppose that with probability $\theta\in(0,1)$ there is a negative demand shock. Each airline receives a private signal $\theta_i$ of the actual $\theta$ and publicly announces its $\theta_i$ during its earnings call, and airlines then base their decisions on all the announced $\theta$'s. So, airlines reduce capacity when all signals are unfavorable compared to when only one firm received a negative signal because of the correlation in their strategies induced by information sharing.\footnote{  This alternative model makes a stronger assumption---airlines cannot misrepresent their information. Under our cheap-talk interpretation, however, it is moot whether or not a message is truthful.}

This alternative model assumes that airlines always have an incentive to share their information about aggregate demand. \citemain{Clarke1983}, \citemain{Galor1985}, and \citemain{Li1985}, however, show that firms do not have an incentive to share their private information about market demand with others unless, as \citemain{Clarke1983} shows, they can use that information to collude.\footnote{ For more on the role of information-sharing on collusion see \citepmain{Vives2008, SugayaWolitzky2018}.}

To verify the validity of the alternative model, we test its implication that absent its signal about low demand airline $j$ would still reduce capacity in the presence of a strong, aggregate signal from others. To that end, we estimate the effect of ``everyone except airline $j$ talking"  on $j$'s capacity choice next quarter. Let $\texttt{Capacity}\texttt{-Discipline-(not}-j)_{m,t}\in\{0,1\}$ be a dummy variable equal to 1 if the market $m$ in period $t$ is talk eligible and if every legacy carrier serving $m$ except airline $j$ discusses capacity discipline, and $0$ otherwise. Then we estimate \cref{eq:model_capdis} after replacing $\texttt{Capacity-Discipline}_{m,t}$ with $\texttt{Capacity-Discipline-(not}-j)_{m,t}$ and present the results in column (5) of \cref{table:robustness}. 
We find that even when everyone else except $j$ is communicating, it does not affect $j$'s capacity. Although this ``no-effect" result is inconsistent with the information-sharing model, it is consistent with the allegation that legacy carriers communicate to coordinate capacity reductions. 

\subsection{Conditional Exogeneity}\label{sec:iv_cond_exog} 

Although we employ a rich set of fixed effects and other covariates as control variables, it is still desirable to explore the possibility that our finding is driven by a missing communication-related variable positively correlated with capacity discipline and negatively correlated with offered seats. To this end, we propose to run a diagnostic test \`a la \citemain{WhiteChalak2010}.

We can explain this approach using an example. 
Suppose we define an additional communication variable equal to 1 whenever all legacy airlines use the word \emph{``stable"}, and zero otherwise. Furthermore, suppose that the occurrence of \emph{``stable"} is positively correlated with, and occurs as frequently as, the discussion of \emph{``capacity discipline.''} 
Then, under this diagnosis, we verify that adding this new dummy variable that captures the discussion of \emph{``stable"} as an additional regressor in \cref{eq:model_capdis} neither affects the estimated relationship between capacity discipline and offered seats nor is it negatively correlated with offered seats. 

Although intuitive, to implement this diagnostic test, we have first to find all relevant tokens (e.g., \emph{``stable"}). Given the large amount of text data we have, it is a nontrivial task to find such tokens objectively. To do so, we use methods from computational linguistics to search our entire text and identify tokens or keywords that (i) are ``close'' in terms of context to the discussion of capacity discipline, and (ii) occur approximately as frequently as ``capacity discipline." Then, for each token, we define a dummy variable $Z_{m,t}$ that is equal to 1 only if all legacy carriers in talk-eligible market $m$ use it in period $t$ and include it as an additional regressor in \cref{eq:model_capdis}. Then, we test if the estimated coefficient for each $Z_{m,t}$ is statistically negative or not, and verify whether the coefficient of capacity discipline changes with the introduction of $Z_{m,t}$. 

To construct such a set of tokens, we identify three tokens that are essential to the concept of capacity discipline: ``capacity discipline,'' ``demand,'' and ``gdp.'' Then, we use the \texttt{word2vec} model from computational linguistics \citepmain{MikolovChenCorradoDean2013} to determine other tokens that are close to these three tokens, using a distance metric that we define shortly below.\footnote{ The \texttt{word2vec} model was developed at Google in 2013 \citepmain{MikolovChenCorradoDean2013} to analyze text data. For an intuitive and accessible explanation, see \citemain{GoldbergLevy2014}. We use the \texttt{gensim} implementation of the \texttt{word2vec} model \citepmain{RehurekSijka2010}.} \texttt{word2vec} allows us to be objective in determining the tokens.

Broadly, the \texttt{word2vec} model is a neural network that maps each unique token we observe in the earnings call transcripts to an $N$-dimensional vector space (in our analysis, $N=300$) in such a way as to preserve the contextual relationships between the tokens. The vector representation of each token is such that contextually similar tokens are located ``close'' to each other, and tokens that are dissimilar are located ``far'' from each other. This sense of ``closeness'' reflects the likelihood that the given tokens appear near each other in the earnings call transcripts. Thus, if ``discipline'' and ``stable'' are close, then the discussion of one term in an earnings call is likely given a  discussion of the other. We directly train the \texttt{word2vec} model using our transcript data, so the derived relationships between words are specific to the context of airlines' earnings calls, as opposed to a more general context. For example, if airline executives use the word ``discipline'' in a contextually different manner than used in more general conversation or writing, our model will account for that.

To measure the similarity of two tokens in the \texttt{word2vec} vector space, we use a commonly used metric called the \emph{cosine similarity metric.} This metric is equal to the cosine of the angle between the vector representation of the two tokens \citemain{Singhal2001}, such that for any two normalized vectors associated with two tokens, $k$, and $\ell$, this measure of similarity is 
\begin{equation*}\label{eq:d-cos}
	d^{\texttt{cos}}( \ell, k) = \frac{k^T \ell}{||k||\cdot||\ell||},
\end{equation*} 
where $||\cdot||$ is the $L^{2}$ norm. When two vectors are the same, cosine similarity is 1, and when they are independent (i.e., perpendicular to each other), it is 0.\footnote{   Note that the cosine metric is a measure of orientation and not magnitude. This metric is appropriate in our cases, as we are interested in comparing the contextual meaning of the words, not in comparing the frequency of the words.}

\begin{figure}\centering \caption{Example of Token Selection Process}
\label{fig:placebo_token_selection}
\includegraphics[scale=0.5]{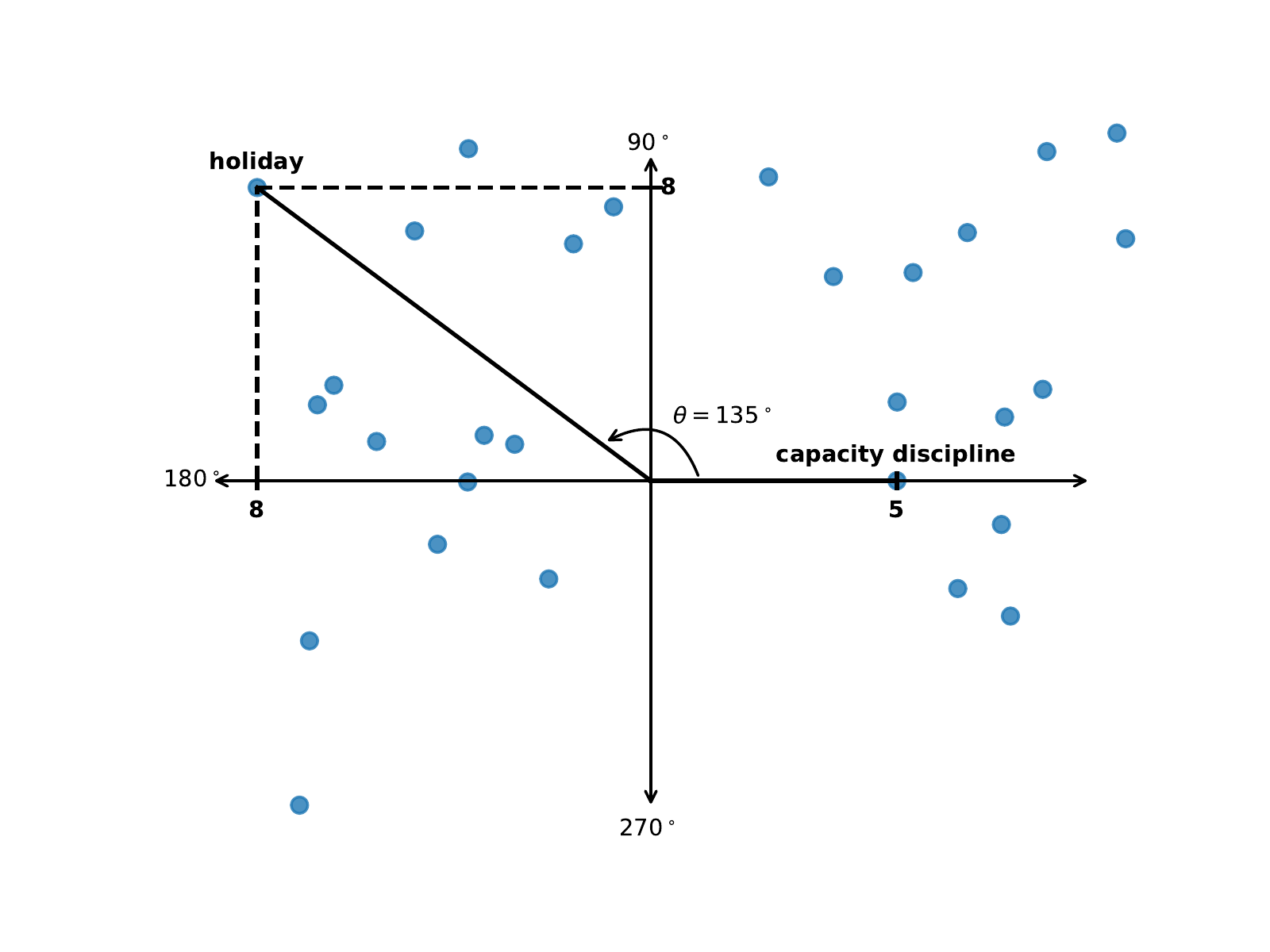}
\subcaption*{\footnotesize Notes. A schematic illustration of a hypothetical \texttt{word2vec} model. Tokens are mapped to a vector space, such that the cosine of the angle between two tokens represents the level of ``similarity'' between those tokens. In the case above, ``holiday'' is seen to be very dissimilar to ``capacity discipline.''}
\end{figure}

To understand our use of cosine similarity, consider \cref{fig:placebo_token_selection}, which displays a hypothetical example of training the \texttt{word2vec} model in a 2-dimensional space. The \texttt{word2vec} model maps all of the tokens in our vocabulary to this space. For example, the token ``capacity discipline'' is represented by the vector $(5,0)$, and the token ``holiday'' is represented by the vector $(-8,8)$. Our measure of similarity between these two tokens is the cosine of the angle between these two vectors, $\theta = 135^\circ$, so $d^{cos}(\texttt{holiday}, \texttt{capacity discipline}) = -0.707$, and thus ``holiday'' is very dissimilar to ``capacity discipline.''

For each of these tokens $k\in\{\texttt{capacity discipline, demand, gdp}\}$, we define the set: 
\begin{equation*} 
L_k(\underline{d}, \overline{d}) = \left\{\ell \in L : \underline{d} \le d^{\texttt{cos}}(\ell, k) \le \overline{d} \right\}, 
\end{equation*} 
where $L$ is the set of all tokens. 
To satisfy the second criterion, we restrict the token to be such that at least 50\% of the time it appears in the same report as these three keywords. 

In \cref{tab.WhiteChalak}, we present all the tokens that satisfy the above two criteria. For each token, we define $Z_{m,t}$ as we did for $\texttt{Capacity-Discipline}_{m,t}$ and use it as an additional regressor in \cref{eq:model_capdis}. The estimated coefficients for the tokens are in the first row, with the estimated coefficient for $\texttt{Capacity-Discipline}_{m,t}$ in the second row. As we can see, five out of six a tokens have no relationship with log seats, and even then the coefficient of  ``domestically" is positive which shows that, if anything, our results understate the true relationship between the discussion of capacity discipline and capacity. What is also reassuring is that for all the tokens, the estimates for \texttt{Capacity-Discipline} are stable, with estimates close to our primary result of $-2.02\%$.

\begin{table}[t!] \centering \caption{Estimates for Conditional Exogeneity}\label{tab.WhiteChalak}
\begin{threeparttable} \scalebox{0.85}{
{
\def\sym#1{\ifmmode^{#1}\else\(^{#1}\)\fi}
\begin{tabular}{l*{6}{c}}
\toprule
                    &\multicolumn{1}{c}{(1)}&\multicolumn{1}{c}{(2)}&\multicolumn{1}{c}{(3)}&\multicolumn{1}{c}{(4)}&\multicolumn{1}{c}{(5)}&\multicolumn{1}{c}{(6)}\\
                    &\multicolumn{1}{c}{slow}&\multicolumn{1}{c}{weakness}&\multicolumn{1}{c}{domestically}&\multicolumn{1}{c}{internationally}&\multicolumn{1}{c}{stable}&\multicolumn{1}{c}{pace}\\
\midrule
Z Token             &     -0.0025&      0.0016&      0.0187&      0.0030&      0.0027&      0.0052\\
                    &    (0.0063)&    (0.0061)&    (0.0068)&    (0.0054)&    (0.0099)&    (0.0076)\\
Capacity Discipline &     -0.0200&     -0.0202&     -0.0187&     -0.0202&     -0.0204&     -0.0207\\
                    &    (0.0059)&    (0.0060)&    (0.0059)&    (0.0059)&    (0.0060)&    (0.0059)\\
\midrule
N                   &     841,991&     841,991&     841,991&     841,991&     841,991&     841,991\\
\bottomrule
\end{tabular}
}

}
\caption*{\footnotesize Notes. Estimation results from including new tokens an additional regressors in \cref{eq:model_capdis}. The table shows the coefficient estimates for each token, and for \texttt{Capacity-Discipline}. We report semi-elasticities (see \cref{footnote:correction}), with standard errors clustered at the bi-directional market level in parentheses. Other control variables included in all regressions, but whose coefficients are not reported are \texttt{Talk-Eligible}, \texttt{Monopoly},  \texttt{MissingReport}, interactions of the \texttt{MissingReport} indicator with \texttt{Talk-Eligible} and \texttt{Monopoly}, origin- and destination-airport annual time trends, carrier-year-quarter fixed effects, and carrier-market fixed effects.
}
\end{threeparttable} \end{table}

\subsection{Additional Robustness Exercises \label{section:robustness}}
In addition to the work described above, we conduct two additional robustness exercises.
For brevity, we present these results in Appendices \ref{sec:cf} and \ref{sec:city-pair}. 

First, in Appendix \ref{sec:cf}, we consider the possibility that market structure can be endogenous because a factor that affects capacity decisions can also affect airlines' decisions to serve a market. If a market structure is endogenous, then \texttt{Capacity-Discipline} will be endogenous as well. To address this, we use a control function approach, where the excluded variables are functions of the geographical distances between a market's endpoints and each carrier's closest hub, which we define as an airport with ``sufficiently" many connections.

The identification assumption is that an airport's distance to the airline's nearest hub is a proxy for entry cost and is therefore correlated with the market structure, but is less likely to be directly correlated with capacity decisions \citepmain{CilibertoTamer2009}. In other words, this approach leverages a timing assumption, namely, that unobservables that affect an airline's network are not contemporaneously correlated with the unobservables that affect the carrier's capacity decisions.  Additionally, the results in Appendix \ref{sec:cf} help to validate our specifications that use carrier-market-structure fixed effects, as the carrier-market-structure fixed effects would violate the strict conditional exogeneity assumption if the market structure is endogenous, which results in biased estimates. 

Second, throughout this paper, we have defined markets as origin and destination airport pairs, an approach commonly used in the literature. A second approach would be to define markets as a directional pair of cities, as discussed in detail in \citemain{BruecknerLeeSinger2014}. In Appendix \ref{sec:city-pair}, we define markets using the city-pair approach and re-estimate our primary specification. Under this approach to defining markets, we fail to find evidence of a relationship between communication and capacity choices. However, this appears to be due to the two three-airport cities in our sample: Washington D.C. and New York City, and excluding them produce results consistent with our primary findings.

\section{Conclusion}\label{section:conclusion}
In this paper, we investigate whether legacy airlines use public communication to sustain cooperation in offering fewer seats in a market. We maintain that airlines communicated, with each other, whenever all legacy carriers serving a market talked about capacity discipline in their earnings calls. Using natural language processing methods, we converted quarterly earnings call transcripts into numeric data to measure communication among legacy carriers. Our estimate is consistent with the allegation that legacy carriers who communicate about ``capacity discipline" offer 2\% fewer seats, on average, across markets and time.
 
 Even though we do not estimate the social value of communication, our estimates suggest that the carriers' capacity reductions are economically significant and most likely harm consumers because (i) we fail to find evidence that the crowding of flight departure times changed; and (ii) simultaneous communication is positively associated with average fares. 
 While we find that these estimates are consistent with anticompetitive behavior, we are aware that communication is not exogenous, and so we have to exercise caution in interpreting these estimation results as proof of collusion.
  
We address various threats to the identification of our primary model. First, while our estimated reduction in capacity after carriers discuss capacity discipline is consistent with airlines coordinating, we do not find it consistent with an alternative hypothesis that earnings calls are serving their intended purpose of making markets more transparent. We also verify that the way we have defined communication in our model is consistent with conditional exogeneity, and finally, we use a control function approach to confirm that our estimates are not affected by endogenous market structure. Thus, we cannot rule out the possibility that public communication allows legacy airlines to coordinate. 

Our finding is relevant for the current policy debate about the social value of information and the correct
response to increasing information about firms in social media and increasing market concentration
across industries. We have provided evidence that in the airline industry, the SEC's transparency regulations are at odds with antitrust laws---a fact that  policymakers should be cognizant of. While the value of public quarterly earnings calls remains debatable, economists and policymakers view the public disclosure of information through these calls as beneficial for investors. At the same time, the competitive effects of this increased transparency are theoretically ambiguous and under-studied. We contribute to this literature and hope that this paper will spur further empirical research on this topic.

While, in some cases, communication helps in equilibrium selection, its broader implications for welfare are unknown. For instance, to determine if a public communication channel is anticompetitive, one must understand how the coordination mechanism depends on the nature of communication. While we find results consistent with the alleged claim that the communication channel enables anticompetitive behavior in the airline industry, there are still many compelling research questions about how these results came to be and the extent to which these results generalize to other industries and methods of communication that remain unanswered. Answers to these questions will help design laws related to public communication and antitrust policy.

In our context of airlines, these questions require the estimation of a flexible oligopoly model, where firms can choose capacity and prices, whether to collude or compete and where strategic behavior can be influenced by public communication. 
As we mentioned earlier, one approach could consist of developing and estimating a model that incorporates both prices and capacity decisions in the airline industry, in the vein of \citemain{KrepsScheinkman1983}, but with differentiated products, and extend it to allow collusion \citepmain{BrockScheinkman1985, BenoitKrishna1987, DavidsonDeneckere1990} with communication. An even more ambitious step would be to allow consumers to care about departures \`{a} la Salop/Vickrey circular city model of \citemain{GuptaLaiPalSarkarYu2004}.
While these models have been studied in isolation, their interactions pose challenges that have not yet been explored and we leave that for future research.

\clearpage \newpage \addcontentsline{toc}{chapter}{Bibliography}
\bibliographystylemain{aea}
\bibliographymain{bibliography}

%%%%%%%%%%%%%%%%%%%%%%%%%%%%%%%%%%%%%%%%

\newpage 
\setcounter{table}{0}
\renewcommand{\thetable}{\Alph{section}.\arabic{table}}
\setcounter{figure}{0}
\renewcommand{\thefigure}{\Alph{section}.\arabic{figure}}
\setcounter{equation}{0}
\renewcommand{\theequation}{\Alph{section}.\arabic{equation}}

\begin{appendices}
\setcounter{table}{0}
\renewcommand{\thetable}{\Alph{section}.\arabic{table}}
\setcounter{figure}{0}
\renewcommand{\thefigure}{\Alph{section}.\arabic{figure}}

\crefalias{section}{appsec}

\setcounter{table}{0}
\renewcommand{\thetable}{\Alph{section}.\arabic{table}}
\setcounter{figure}{0}
\renewcommand{\thefigure}{\Alph{section}.\arabic{figure}}

\section{Control Function Approach}\label{sec:cf}\label{sec:cf_main}

In this section, we present results from using a control function approach to estimate our model.

Our treatment, $\texttt{Capacity-Discipline}_{m,t}$, is the product of $\texttt{Talk-Eligible}_{m,t}$ and whether all of the legacy carriers in $m$ discussed capacity discipline in their most recent earnings calls. By construction, $\texttt{Talk-Eligible}_{m,t}$ is a function of the market structure (the set of airlines who serve market $m$ in month $t$). An airline's decision to serve $m$, among other factors, will depend on the cost of serving it, which is unobserved and might not be captured by the fixed effects. So it is possible that $\texttt{Talk-Eligible}_{m,t}$ is endogenous, which in turn means $\texttt{Capacity-Discipline}_{m,t}$ would be endogenous too. And because $\texttt{Talk-Eligible}_{m,t}$, and hence $\texttt{Capacity-Discipline}_{m,t}$, are negatively correlated with the cost of serving $m$ in $t$, our estimator in \cref{eq:model_capdis} might exaggerate the negative effect of communication on capacity.
 
Finding an IV for our regression is not a simple task because decisions across markets are interconnected in a network industry. For example, \cite{HendricksPiccioneTan1999} consider a one-shot two-stage model where two carriers incur fixed costs and simultaneously choose their networks and compete (Bertrand or Cournot) for passengers. 
Still, we believe that the endogeneity of market structure could bias our results, and so we propose an instrumental variable that exploits a plausible timing assumption. Leveraging a timing assumption in this way is common in empirical studies of market competition; see, for example, \cite{OlleyPakes1999} and \cite{Eizenberg2014}. 

In particular, we propose to use a measure of the distance between a market's endpoints and the carriers' closest hubs, henceforth, ``hub-distances,'' as an instrumental variable for market structure. The distance of a market's endpoints to a carrier's closest hub is a proxy for the fixed cost that a carrier has to face to serve that market \citepmain{CilibertoTamer2009}. This is the direct effect of the distance on an airline's decision to serve a market. Distances to the hubs also indirectly affect the market structure through competition: An airline's probability of serving a market should increase with its competitors' distances.

Conditional on including the distance between the origin and destination airport, which is captured by the carrier-market fixed effects in our model, hub-distance should not affect consumer demand and the carriers' variable costs. We are aware that, for a given network structure of the industry, the distance to hubs might correlate with capacity decisions, but we believe that if the relationship exists, it is weaker than the one between the distances from the hubs and the entry decision. 
Indeed, the variable hub-distance is not included in the standard structural models of demand and supply for the airline industry, see \cite{BerryJia2010}. Moreover, the fact that we measure the impact of communication on market-level capacity choices and not on the aggregate capacities further suggests that hub-distance is uncorrelated with the capacity choice.

Finally, as mentioned above, our instruments rely on a plausible timing assumption, namely that the unobservables that affect the development of an airline network are not \textit{contemporaneously} correlated with the unobservables that affect prices and capacity decisions. This assumption relies on a crucial institutional feature of the airline industry, whereby network service is fundamentally dependent on the ability of airlines to enplane and deplane travelers at airports, and they can only do this if they have access to gates. As discussed in \cite{CilibertoWilliams2014}, ``a substantial majority of gates are leased on an exclusive or preferential basis, and for many years.'' In addition, \citeauthor{CilibertoWilliams2014} note that ``it is difficult to adjust access to airport facilities in response to unexpected changes in demand and costs.'' This institutional feature of the airline industry is particularly true for an airline's hubs. Therefore, the development of an airline network, with its determination of its hub-and-spoke structure, is considerably slower than an airline's ability to enter and exit markets, and to change capacities and prices. Despite this, we are aware that there might be persistent components of the unobservables, but we maintain that those are captured by the market-carrier and the carrier-year-quarter fixed effects.

To measure the role of airline networks as determinants of market structure we proceed as follows. First, for each airline, we compute the air-distance of an airport to the airline's ``hubs" (which are defined based on connectedness of the time-varying network of markets served by an airline, defined shortly below).\footnote{ The concept of connectedness is from the theoretical literature on networks. See \cref{sec:cf_iv} for additional details on the calculation of the set of hubs for each airline.} Data on the distances between airports are from the data set \emph{Aviation Support Tables: Master Coordinate}, available from the National Transportation Library.  
Then for each carrier $j$, market $m$, and month $t$ in our sample, we calculate that carrier's hub-distance $D_{j,m,t}$ as the sum of the distance from the origin airport to the carrier's nearest hub, and the distance from the destination airport to the carrier's nearest hub. We use these hub-distances as instrumental variables for $\texttt{Talk-Eligible}$ and, in turn, for $\texttt{Capacity-Discipline}$.\footnote{We thank Mar Reguant for suggesting this approach, that when an endogenous variable is a interaction then we can use one of the two variables as an instrument for the product, and if that variable is also endogenous then an instrument for that variable will still be a valid instrument for the product. It is similar to the approach used in \cite{FabraReguant2014}. Our approach also controls for an (unlikely) event that legacy carriers discussing capacity correlates with the unobserved cost of serving a market, as long as that event is uncorrelated with the instrumental variable.}

\begin{figure}[h!]\centering \caption{Histogram of the Standard Deviation of Hub-distances across Carrier-Markets}
\label{figure:distance_distribution}
\begin{subfigure}[t]{0.45\textwidth}\centering
\includegraphics[scale=0.4]{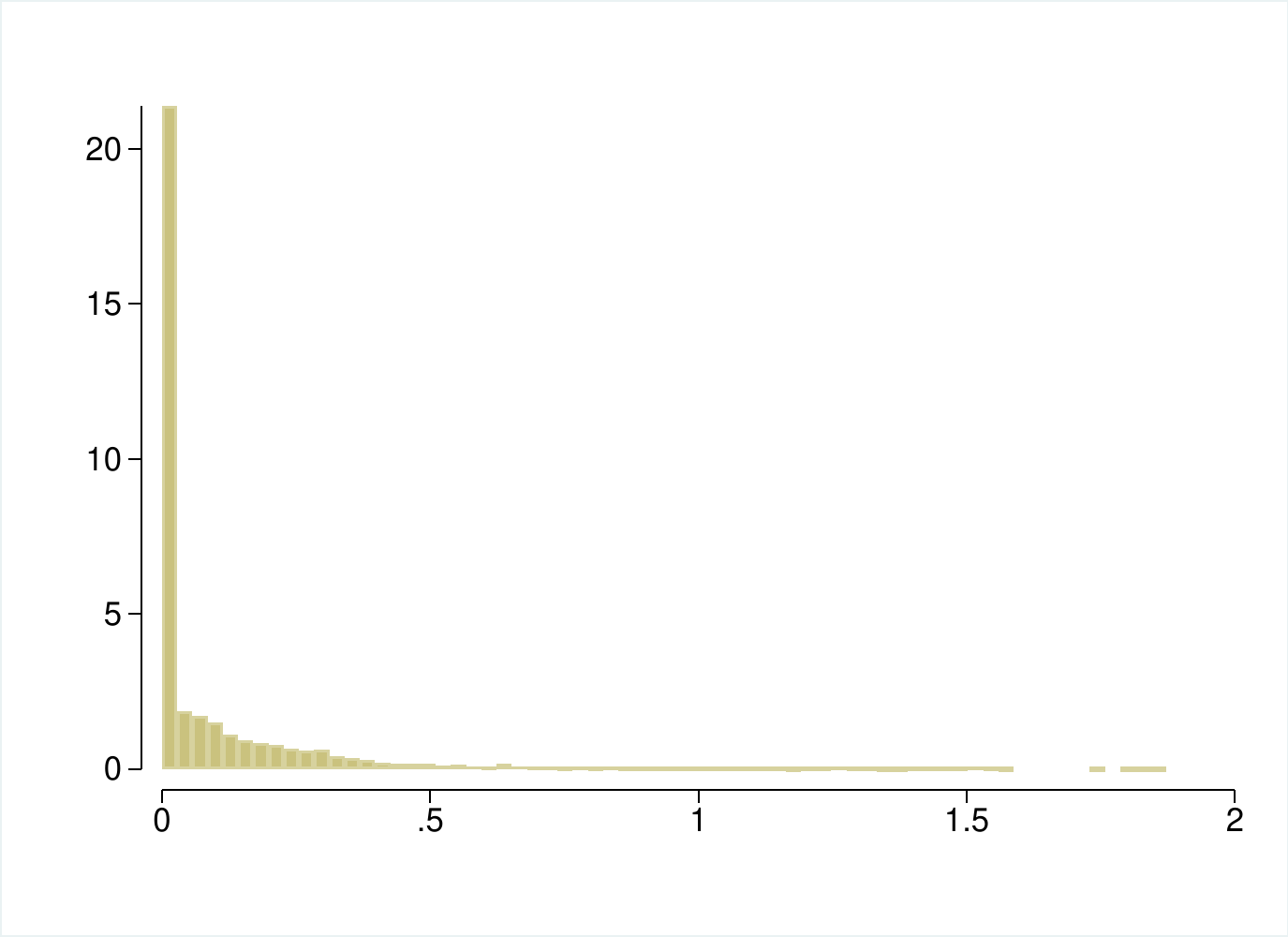}
\caption{All Values}\label{figure:distance_distributionall}
\end{subfigure}
\begin{subfigure}[t]{0.45\textwidth}\centering
\includegraphics[scale=0.4]{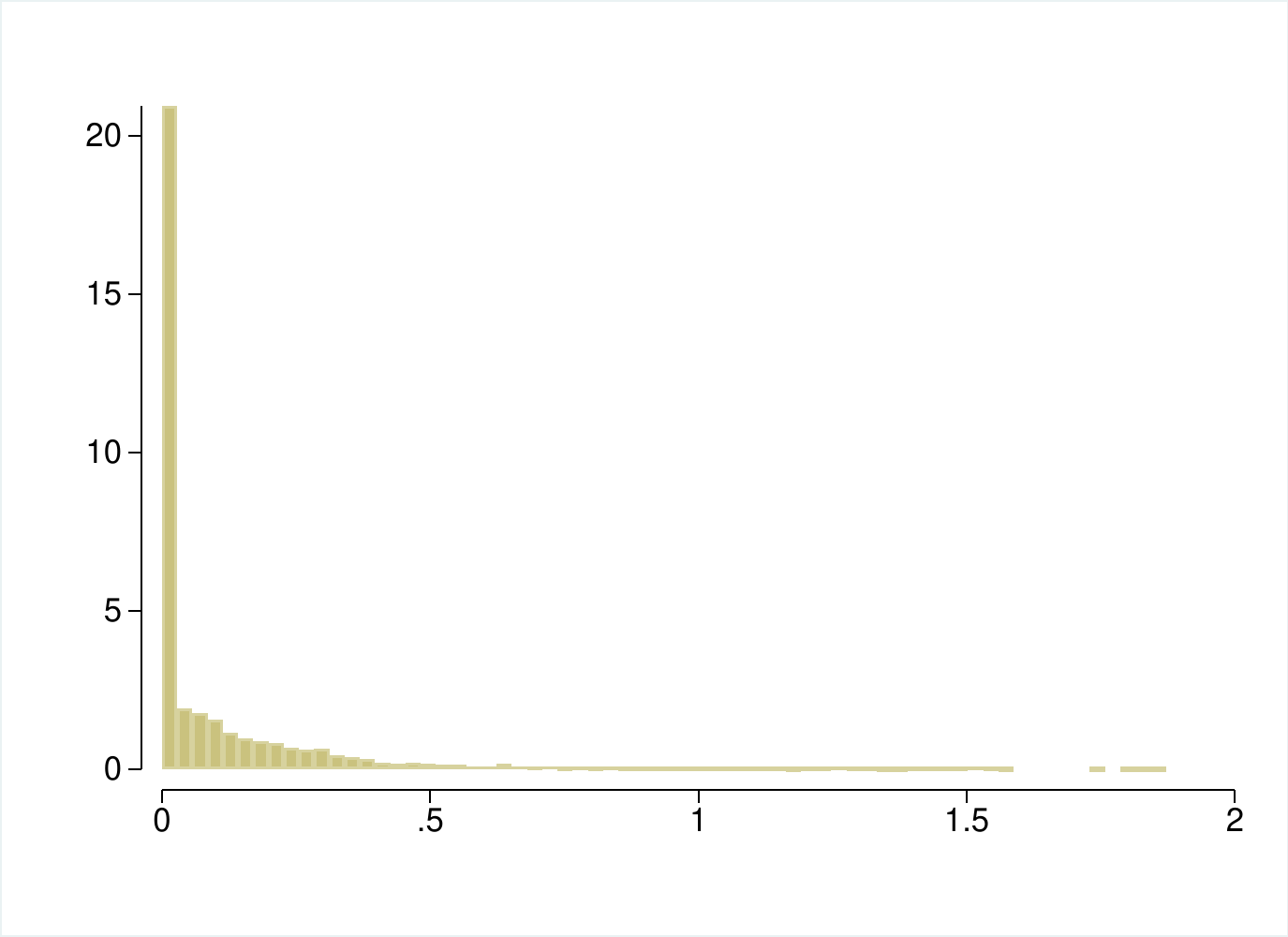}
\caption{Positive Values}\label{figure:distance_distributionpositive}
\end{subfigure}
\subcaption*{\footnotesize Notes. Observations constructed by calculating the standard deviation of the hub-distance for each carrier-market. Hub-distance is measured in thousands of miles. Panel (a) includes all carrier-market observations, and panel (b) only includes carrier-market observations where the value is non-zero.}
\end{figure}
\begin{table}[t!]
\centering
\caption{Summary Statistic of Hub-distances by Carriers}
\label{table:distancesummary}
\begin{tabular}{llllll}\toprule
	&\textbf{Mean}&\textbf{Sd}&\textbf{Median}&\textbf{N} \\
\midrule
\textbf{Airline}&&&& \\
AA&1.274&0.629&1.192&613,673 \\
AS&3.550&1.113&3.801&613,673 \\
CO&1.315&0.770&1.150&613,673 \\
DL&1.059&0.506&0.987&613,673 \\
LCC&0.956&0.608&0.837&613,673 \\
NW&1.258&0.711&1.054&613,673 \\
UA&1.089&0.543&1.038&613,673 \\
US&1.215&0.745&1.068&613,673 \\
\midrule
\textbf{Total}&1.496&1.120&1.146&4,909,384 \\
\bottomrule
\end{tabular}
\subcaption*{\footnotesize Notes. Each row displays the mean, standard deviation, median and number of observations of air-distances to closest hubs for a carrier. Distances are measured in thousands of miles. LCC is the average of distances for all LCCs.}	
\end{table}
In \cref{figure:distance_distribution} we display the histograms for the within carrier-market variances of  these distances, measured in thousands of miles. \cref{figure:distance_distributionall} displays the entire sample while \cref{figure:distance_distributionpositive} restricts the sample to only those with positive variance in distances. Both these figures and table show that there is substantial variation in distances. We also present the summary statistics of these distances by carriers in \cref{table:distancesummary}.

\begin{table}[t!]\centering
\caption{First-stage Regression for Control Function Approach: Communication and Available Seats}\label{tab:cf_first_stage}
\begin{threeparttable} \scalebox{0.9}{
{
\def\sym#1{\ifmmode^{#1}\else\(^{#1}\)\fi}
\begin{tabular}{l*{1}{c}}
\toprule
                    &\multicolumn{1}{c}{(1)}\\
                    &\multicolumn{1}{c}{Talk Eligible}\\
\midrule
AA Distance         &     -0.0195\\
                    &    (0.0014)\\
CO Distance         &      0.0088\\
                    &    (0.0009)\\
DL Distance         &     -0.0281\\
                    &    (0.0016)\\
LCC Distance        &     -0.0196\\
                    &    (0.0007)\\
NW Distance         &     -0.0051\\
                    &    (0.0012)\\
UA Distance         &     -0.0011\\
                    &    (0.0013)\\
US Distance         &     -0.0158\\
                    &    (0.0006)\\
\midrule
F-statistic (instruments)&    265.9674\\
N                   &     613,673\\
\bottomrule
\end{tabular}
}

}
\caption*{\footnotesize Notes. Observations are at the market-month level. Bootstrapped standard errors clustered at the bi-directional market level are reported in parentheses. Other control variables included in the regression, but whose coefficients are not reported are \texttt{Monopoly},  \texttt{MissingReport}, the interaction interaction of the \texttt{MissingReport} and \texttt{Monopoly} variables, origin- and destination-airport annual time trends, year-quarter fixed effects, and market fixed effects.
}
\end{threeparttable}
\end{table}

Using the calculated hub-distances, we employ a control function approach to estimate the effect of communication on capacity  \citep{ImbensWooldridge2007}. In the first-stage, we estimate 
\begin{equation}\label{eq:cf_first_stage}
	\begin{aligned}
		\texttt{Talk-}&\texttt{Eligible}_{m,t}= \sum_{j \in J^{CF}} \sigma_j D_{j,m,t} \\
		&\qquad+ \alpha_{0}\times \texttt{Monopoly}_{m,t}+\alpha_{1} \times \texttt{MissingReport}_{m,t}\\
		&\qquad+ \alpha_{2}\times \texttt{Monopoly}_{m,t} \times \texttt{MissingReport}_{m,t}\\
		&\qquad+ \mu_{m} + \mu_{yr,q} + \gamma_{origin,t} + \gamma_{destination,t} + r_{m,t},
	\end{aligned}
\end{equation}
where $J^{CF}$ is the set of legacy carriers, and an aggregated LCC carrier. We aggregate the low-cost carriers by setting $D_{LCC, m, t}$ to the shortest hub-distance for all of the LCCs for market $m$ in month $t$. We present the results of estimating \cref{eq:cf_first_stage} in \cref{tab:cf_first_stage}. Having estimated \cref{eq:cf_first_stage} we recover the residuals $\hat{r}_{m,t}$. Then, in the second step, we re-estimate the parameters in \cref{eq:model_capdis} with $\hat{r}_{m,t}$ as an additional covariate. We present the second-stage results in column (1) of \cref{tab:cf}, and replicate our primary results in column (2) to facilitate comparison. We can see that when legacy carriers communicate, they reduce their capacity by $2.01\%$. Thus, we still find strong evidence that airlines use earnings calls to coordinate in reducing their capacities.

\begin{table}[t!]\centering
\caption{Control Function Approach: Communication and Available Seats}\label{tab:cf}
\begin{threeparttable} \scalebox{0.9}{
{
\def\sym#1{\ifmmode^{#1}\else\(^{#1}\)\fi}
\begin{tabular}{l*{2}{c}}
\toprule
                    &\multicolumn{1}{c}{(1)}&\multicolumn{1}{c}{(2)}\\
                    &\multicolumn{1}{c}{Log Seats}&\multicolumn{1}{c}{Log Seats}\\
\midrule
Capacity-Discipline &     -0.0201&     -0.0202\\
                    &    (0.0060)&    (0.0060)\\
Talk Eligible       &      0.1478&     -0.0415\\
                    &    (0.2331)&    (0.0150)\\
Residual            &     -0.1904&            \\
                    &    (0.2349)&            \\
\midrule
N                   &     841,353&     841,991\\
\bottomrule
\end{tabular}
}

}
\caption*{\footnotesize Notes. We report semi-elasticities (see \cref{footnote:correction}), with bootstrapped standard errors clustered at the bi-directional market level in parentheses. Column (1) reports the results of the control function approach, and column (2) replicates our primary estimates of \cref{eq:model_capdis} to facilitate comparison. Other control variables included in all regressions, but whose coefficients are not reported are \texttt{Talk-Eligible}, \texttt{Monopoly},  \texttt{MissingReport}, interactions of the \texttt{MissingReport} indicator with \texttt{Talk-Eligible} and \texttt{Monopoly}, origin- and destination-airport annual time trends, carrier-year-quarter fixed effects, and carrier-market fixed effects.
}
\end{threeparttable}
\end{table}

\subsection{Determination of Airline Hubs}\label{sec:cf_iv}
In this section, we explain how we determine hubs for each airline, and provide evidence of variations in our instruments. To identify hubs over time, we follow \cite{CilibertoCookWilliams2018}. They show that using the shortest path between two airports to determine the \emph{betweenness centrality} measure identifies the hub airports well. 

To illustrate this measure of centrality, consider Figure \ref{fig:network}, which displays a network of airports served by an airline. Betweenness centrality for CHO measures the number of times CHO is the shortest connection between two other airports. In this example, CHO is never in the shortest path between any two airports, so the betweenness centrality for CHO is zero. Similarly, the betweenness centrality for PHX is also zero. However, DFW will have higher betweenness centrality because it is in a stop of multiple airports, like PHX and SFC. Similarly, the betweenness centrality for CLT and LAX will be high. 

\begin{figure}[t!]
\begin{center}
\caption{ Network for an Airline\label{fig:network}}
\begin{tikzpicture}[auto, thick]
 \foreach \place/\name in {{(0,-1)/a}, {(2,0)/b}, {(2,2)/c}, {(0,2)/d},
      {(-2,0)/e}}
  \node[superpeers] (\name) at \place {};
 \foreach \source/\dest in {a/b, a/c, a/d, b/c, c/d,a/e,d/e}
  \path (\source) edge (\dest);
 \foreach \pos/\i in {above left of/1, left of/2, below left of/3}
  \node[peers, \pos = e] (e\i) {};
  \foreach \speer/\peer in {e/e1,e/e2,e/e3}
  \path (\speer) edge (\peer);
  \foreach \pos/\i in {above right of/1, right of/2, below right of/3}
  \node[peers, \pos =b ] (b\i) {};
  \foreach \speer/\peer in {b/b1,b/b2,b/b3}
  \path (\speer) edge (\peer);
  \node[peers, above of=d] (d1){};
  \path (d) edge (d1);
  \foreach \pos/\i in {below left of/1, below of/2}
  \node[peers, \pos =a ] (a\i) {};
  \foreach \speer/\peer in {a/a1,a/a2}
  \path (\speer) edge (\peer);
  %%%%%%%%
  % Legends
  % for large airports
  \node at (0,-1) {\tiny{DFW}};
  \node at (2,0) {\tiny{CLT}};
  \node at (2,2) {\tiny{JFK}};
  \node at (0,2) {\tiny{ORD}};
  \node at (-2,0) {\tiny{LAX}};
  % for small airports
  \node at (3.45,0) {\tiny{CHO}};
  \node at (0,-2.3) {\tiny{PHX}};
  \node at (-3.45,0) {\tiny{SFO}};
\end{tikzpicture}
\end{center}
\caption*{\footnotesize Notes. A schematic representation of airports-network served by an airline.}
\end{figure}

Formally, the betweenness measure for an airport $k$, for airline $j$ is
\begin{eqnarray*}
B_k^j \quad:=\quad \sum_{\ell\neq \ell', k\not\in\{\ell, \ell'\}}\frac{1}{(N_j-1)(N_j-2)} \frac{P_k^j(\ell, \ell')}{P^j(\ell,\ell')},
\end{eqnarray*} 
where $N_j$ is the number of airports served by airline $j$, $P_k^j(\ell,\ell')$ is the number of shortest paths between airports $\ell$ and $\ell'$ with a stop at $k$, and $P^j(\ell, \ell')$ is the total number of shortest paths between $\ell$ and $\ell'$. If there is only one shortest path between $\ell$ and $\ell'$, then the ratio is 1, and if there are multiple paths, then this measure gives equal weight to each path. The measure is rescaled by dividing through by the number of pairs of nodes not including $k$, so that $B_{k}^{j}\in[0,1]$.
Using this measure of betweenness centrality, for every airline $j$ and every period $t$ we choose the airports with the betweenness centrality that is at least $0.1$ and denote these airports as $j$'s ``hubs.'' By this definition, the hubs in Figure \ref{fig:network} are $\{DFW, CLT, LAX\}$.

 \setcounter{table}{0}
\renewcommand{\thetable}{\Alph{section}.\arabic{table}}
\setcounter{figure}{0}
\renewcommand{\thefigure}{\Alph{section}.\arabic{figure}}

\section{Market Heterogeneity}\label{sec:market_hetero}

In this section, we consider the role of market sizes and the composition of passengers in determining the relationship between communication and capacity choices.

\subsection{The Role of Market Size}\label{sec:market_size}

First, we explore how airlines' reductions in capacity differ by market size. Carriers' ability to coordinate on capacity can vary by market, depending on the ability of legacy airlines to monitor each other and their markets' contestability. We follow \cite{BerryCarnallSpiller2006} and define market size as the geometric mean of the Core-based statistical area population of the end-point cities. The annual population data are from the U.S. Census Bureau. We define markets with a population larger than the $75^{th}$ percentile of the market population distribution as large, markets with a population in the range of $(25^{th}, 75^{th}]$ percentiles of the population as medium, and those at or below the $25^{th}$ percentile as small markets.\footnote{ When classifying markets as small, medium, or large, we use the average market population over our sample period so that a market's size classification does not change across time. The $25^{th}$ percentile cutoff is 1.27 million people, and the $75^{th}$ percentile cutoff is 3.25 million people.}

\begin{table} \caption{Summary Statistics for Market Size and Business Travel}\label{table:sum-stat_marketTypes_airportpair}
\begin{adjustbox}{width=1\textwidth}
	\begin{tabular}{lllllllllllll}\toprule 
		& \multicolumn{3}{c}{Seats} & \multicolumn{2}{c}{Cap. Discipline} & \multicolumn{2}{c}{Talk Eligible} & \multicolumn{2}{c}{Monopoly Market} & \multicolumn{2}{c}{Missing Report}&\\
\cmidrule(l{.75em}){2-4} \cmidrule(l{.75em}){5-6}\cmidrule(l{.75em}){7-8}\cmidrule(l{.75em}){9-10}\cmidrule(l{.75em}){11-12}
&Mean&SD&Median&Mean&SD&Mean&SD&Mean&SD&Mean&SD&N \\
\midrule
\textbf{Market Size}&&&&&&&&&&&& \\
Small&5,301.915&5,469.491&3,844.000&0.005&0.069&0.027&0.163&0.842&0.365&0.192&0.394&115,293 \\
Medium&9,890.301&9,373.000&7,145.000&0.046&0.209&0.156&0.362&0.588&0.492&0.191&0.393&420,714 \\
Large&16,298.972&14,270.025&11,880.000&0.129&0.336&0.445&0.497&0.309&0.462&0.236&0.425&305,984 \\
\midrule
\textbf{Business Travel}&&&&&&&&&&&& \\
Low Business&11,249.436&11,404.137&7,535.000&0.067&0.250&0.216&0.411&0.445&0.497&0.202&0.402&175,445 \\
Medium Business&12,005.773&12,151.691&7,949.000&0.089&0.285&0.296&0.456&0.461&0.499&0.232&0.422&295,324 \\
High Business&11,605.024&11,495.510&7,883.000&0.061&0.239&0.216&0.412&0.599&0.490&0.218&0.413&149,993 \\
\midrule
\textbf{Total}&11,590.963&11,700.888&7,776.000&0.070&0.256&0.243&0.429&0.521&0.500&0.207&0.405&841,991 \\

		\bottomrule
	\end{tabular}
\end{adjustbox}
\caption*{\footnotesize Notes: Observations are at the carrier-market-month level.}
\end{table}

\cref{table:sum-stat_marketTypes_airportpair} shows that the average number of seats a carrier offers, the likelihood of the treatment $\texttt{Capacity-Discipline}=1$, and the likelihood of $\texttt{Talk Eligible}=1$ are increasing with the size of a market. Perhaps unsurprisingly, the likelihood that a market is a monopoly market decreases with the size of the market. 

To assess if the intensity of coordinated capacity reduction varies with market size, we re-estimate \cref{eq:model_capdis}, interacting $\texttt{Capacity-Discipline}_{m,t}$ with indicators of whether a market is small, medium, or large.\footnote{ Although not reported, we also allow the impacts of $\texttt{Talk-Eligible}$, $\texttt{Monopoly}$, and $\texttt{MissingReport}$ to vary with market size.} We present these estimation results in column (1) of \cref{tab.pop_biz}. We find that communication among legacy carriers is associated with, on average, a $1.55\%$ and $1.40\%$ reduction in seats supplied in smaller and medium markets, respectively, but that the coefficients are imprecisely estimated and, as a result, we cannot reject the null hypothesis that in these two types of markets, communication and the number of seats offered are uncorrelated. However, for the large markets, we find that communication among legacy carriers is associated with a $2.42\%$ reduction in seats supplied. 

\begin{table}[t!]\centering
\caption{Communication and Available Seats: The Role of Market Size and Business Travel}\label{tab.pop_biz}
\begin{threeparttable} \scalebox{0.75}{
{
\def\sym#1{\ifmmode^{#1}\else\(^{#1}\)\fi}
\begin{tabular}{l*{2}{c}}
\toprule
                    &\multicolumn{1}{c}{(1)}&\multicolumn{1}{c}{(2)}\\
                    &\multicolumn{1}{c}{Log Seats}&\multicolumn{1}{c}{Log Seats}\\
\midrule
Capacity Discipline x Small Population&     -0.0146&            \\
                    &    (0.0276)&            \\
Capacity Discipline x Medium Population&     -0.0134&            \\
                    &    (0.0108)&            \\
Capacity Discipline x Large Population&     -0.0251&            \\
                    &    (0.0069)&            \\
Capacity Discipline x Low Business&            &     -0.0087\\
                    &            &    (0.0098)\\
Capacity Discipline x Medium Business&            &     -0.0266\\
                    &            &    (0.0078)\\
Capacity Discipline x High Business&            &     -0.0172\\
                    &            &    (0.0130)\\
\midrule
R-squared           &       0.088&       0.086\\
N                   &     841,991&     620,762\\
\bottomrule
\end{tabular}
}

}
\caption*{\footnotesize Notes. We report semi-elasticities (see \cref{footnote:correction}), with standard errors clustered at the bi-directional market level in parentheses. Other control variables included in all regressions, but whose coefficients are not reported are \texttt{Talk-Eligible}, \texttt{Monopoly},  \texttt{MissingReport}, interactions of the \texttt{MissingReport} indicator with \texttt{Talk-Eligible} and \texttt{Monopoly}. These coefficients are allowed to vary based on the market size or business travel classifiers. Additionally, all regressions include origin- and destination-airport annual time trends, carrier-year-quarter fixed effects, and carrier-market fixed effects.}
\end{threeparttable}
\end{table}

\subsection{The Role of Business Travelers}\label{sec:biz-travelers}

Next, we investigate whether the composition of the market demand in business and leisure travelers is associated with the degree to which carriers respond to communication. Business travelers tend to have a higher willingness to pay for a ticket and have less elastic demand for air travel than leisure travelers. So, all else equal, markets with a relatively high number of business travelers should have higher mark-ups and be more lucrative for coordination.  

We follow \cite{Borenstein2010} and \cite{CilibertoWilliams2014} and use a business index constructed using the 1995 American Travel Survey (ATS). The ATS was conducted by the Bureau of Transportation Statistics (BTS) to obtain information about the long-distance travel of people living in the U.S., and it collected quarterly information related to the characteristics of persons, households, and trips of 100 miles or more for approximately 80,000 American households. We use the survey to compute an index that measures the fraction of passengers traveling for business out of an airport.

We define a market's business travel index as the computed travel index for its origin airport. In classifying markets based on their business travel level, we follow the same approach as in our market size classifications. Low business markets are those with an index value at or below the 25$^{th}$ percentile, medium business markets have an index value in the $(25^{th}, 75^{th}]$ percentiles, and high business markets are those with an index above the $75^{th}$ percentile. The average number of seats offered in a market is relatively constant across our business travel classifications, but coordinated communication is more common in low and medium business markets than in high business markets. Having constructed our business classifications, we estimate a model interacting $\texttt{Capacity-Discipline}_{m,t}$ with indicators for the three levels of business travel.

We present the results from this regression column (2) of \cref{tab.pop_biz}. The last three rows present the estimated relationship between communication and capacity choices in low-business, medium-business, and high-business markets, respectively. As we can see, communication is associated with a 0.09\%, 2.66\%, and 1.72\% decrease in the number of seats offered, respectively, although the estimates for low- and high-business markets are imprecisely estimated.

\section{An Alternative Approach to Defining Markets: City Pairs}\label{sec:city-pair}

In the main paper, we have followed \cite{Borenstein1989, KimSingal1993, BorensteinRose1994, GerardiShapiro2009, CilibertoTamer2009, BerryJia2010, CilibertoWilliams2010}; and \cite{CilibertoWilliams2014}, and defined a market by the origin and
destination airport pairs. An alternative argument maintains that markets are to be defined by the origin and destination \emph{cities},
rather than airports. This alternative market definition has been followed by, among others, \cite{Berry1990, Berry1992, BruecknerSpiller1994, EvansKessides1994}; and \cite{ BambergerCarltonNeumann2004}. 

\begin{table}[h!]\centering
\caption{Summary Statistics for City-Pair Markets}\label{table:sum-stat_marketTypes_city_pair}
\begin{subfigure}{\linewidth}\centering
	\begin{adjustbox}{width=1\textwidth}
	\begin{tabular}{lllllllllllll}\toprule & \multicolumn{3}{c}{Seats} & \multicolumn{2}{c}{Cap. Discipline} & \multicolumn{2}{c}{Talk Eligible} & \multicolumn{2}{c}{Monopoly Market} & \multicolumn{2}{c}{Missing Report}&\\
\cmidrule(l{.75em}){2-4} \cmidrule(l{.75em}){5-6}\cmidrule(l{.75em}){7-8}\cmidrule(l{.75em}){9-10}\cmidrule(l{.75em}){11-12}
&Mean&SD&Median&Mean&SD&Mean&SD&Mean&SD&Mean&SD&N \\
\midrule
\textbf{Carrier Type}&&&&&&&&&&&& \\
Legacy&12,751.265&15,306.754&7,440.000&0.108&0.311&0.395&0.489&0.445&0.497&0.279&0.448&518,609 \\
LCC&11,694.473&11,826.904&8,220.000&0.076&0.265&0.270&0.444&0.295&0.456&0.165&0.372&269,019 \\
\midrule
\textbf{Market Participants}&&&&&&&&&&&& \\
Mixed Market&14,919.775&16,107.449&9,381.000&0.115&0.319&0.415&0.493&0.166&0.372&0.224&0.417&478,473 \\
Legacy Market&8,475.512&9,413.635&5,428.000&0.070&0.256&0.254&0.435&0.746&0.435&0.265&0.442&309,155 \\
\midrule
\textbf{Total}&12,390.312&14,223.136&7,809.000&0.097&0.296&0.352&0.478&0.393&0.488&0.240&0.427&787,628 \\
\bottomrule
	\end{tabular}
	\end{adjustbox}
\end{subfigure}
\caption*{\footnotesize Notes. Table of summary statistic for all key variables. Observations are at the carrier-market-month level for city-pair markets.}
\end{table} 

The city-pair market aggregates possibly more than one airport-pair market. For illustration, consider two flights flying out of Piedmont Triad International Airport (GSO), located in Greensboro, NC, with one flying to O'Hare International Airport (ORD) and the other flying to Midway International Airport (MDW), both located in Chicago, IL. Under the airport-pair market definition, these flights operate in separate markets---the first is in the GSO-ORD market, and the second is in the GSO-MDW market. Under the city-pairs market definition, these flights operate in the same Greensboro to Chicago market.\footnote{ In our empirical analysis, to identify the airports under the city-pair definition, we follow \cite{BruecknerLeeSinger2014}.} In Table \ref{table:sum-stat_marketTypes_city_pair} we present the city-pair analogue of \cref{table:sum-stat}. 

How to define airline markets is of interest in antitrust policies. While the airport-pair approach is often used in academic research on the airline industry, antitrust practitioners use the city-pair approach. Using the city-pair approach leads to larger markets, which, for antitrust purposes, provides a more robust basis for government intervention if there is any evidence of anticompetitive effects.

\begin{table}[t!]
\caption{Communication in Airport- vs. City-pair Markets}\label{table:city-pair-example}
\resizebox{\textwidth}{!}{
\begin{tabular}{ll|ll|ll|ll|ll}\toprule
\multicolumn{2}{c|}{City}&\multicolumn{2}{c|}{Airport}&&&\multicolumn{2}{c}{\texttt{Talk-Eligible}}&\multicolumn{2}{|c}{\texttt{Capacity-Discipline}}\\
 Origin 	& Destination	& Origin 	& Destination	& Carrier 		& Communication 	& Airport-pair	& City-pair	& Airport-pair 	& City-pair\\\midrule
\multirow{4}{*}{Greensboro, NC} & \multirow{4}{*}{Chicago, IL}		& \multirow{2}{*}{GSO}		& \multirow{2}{*}{ORD}			& AA (legacy)	& 1		& 1			& \multirow{4}{*}{1}			& \multirow{2}{*}{1}			& \multirow{4}{*}{0}	 	\\
 			& 				& 			& 				& DL (legacy)	& 1		& 				& 				&				& 			\\
 			& 				& \multirow{2}{*}{GSO}			& \multirow{2}{*}{MDW}				& UA (legacy)	& 0		& \multirow{2}{*}{0}			& 				& \multirow{2}{*}{0}			& 			\\
 			& 				& 			&				& B6 (lcc)		& N/A		& 				& 				&				& 			\\
\bottomrule
\end{tabular}
}\caption*{\footnotesize Notes. Table shows an example that highlight changes in our definition of communication when we move from airport-pair definition to city-pair definition of a market.}	
\end{table}

However, with the city-pair definition, we should expect the effect of communication on the capacity to change. 
As an example, consider \cref{table:city-pair-example}, which lists four flights from Greensboro, NC to Chicago, IL. Under the airport-pair definition of markets, this table presents two markets: GSO-ORD and GSO-MDW. The first, GSO-ORD, is served by two legacy carriers (AA and DL) and is, therefore, a ``talk eligible'' market. Since both carriers talked about capacity discipline, \texttt{Capacity-Discipline} is equal to 1. However, the second market, GSO-MDW, is served by one legacy, which is not discussing capacity discipline, and one low-cost carrier. Since the market is not talk-eligible, \texttt{Capacity-Discipline} equals 0. As can be seen, under the airport-pair approach to defining markets, we have one market where coordinated communication is taking place and one where it is not.

Now consider the city-pair approach to defining markets. Under this approach, the table shows a single market, Greensboro to Chicago, served by four carriers. Three legacy carriers serve the market, so this city-pair market is talk-eligible. However, one of the legacy carriers did not discuss capacity discipline (UA), so \texttt{Capacity-Discipline} is equal to zero. This example shows how the frequency of $\texttt{Capacity-Discipline}_{m,t}=1$ can differ between airport and city markets. Moreover, depending on the relative passenger volume through GSO-ORD and GSO-MDW, we can get a different result. 
If a city has three airports, then the association between communication and capacity will become even more ambiguous and cannot be predicted by looking at what is happening in those three airports individually. Only two cities, Washington, D.C., and New York City, are served by three airports. Thus, the effects of communication on capacity may vary with market definitions.

\begin{table}[t!] \centering \caption{Communication and Available Seats for City-Pair Markets \label{table:city.main}} 
\begin{threeparttable} 
\scalebox{0.8}{
{
\def\sym#1{\ifmmode^{#1}\else\(^{#1}\)\fi}
\begin{tabular}{l*{2}{c}}
\toprule
                    &\multicolumn{1}{c}{(1)}&\multicolumn{1}{c}{(2)}\\
                    &\multicolumn{1}{c}{Log Seats}&\multicolumn{1}{c}{Log Seats}\\
\midrule
Capacity Discipline &      0.0015&     -0.0185\\
                    &    (0.0056)&    (0.0086)\\
Exclude NYC \& DC & No & Yes\\\midrule
R-squared           &       0.090&       0.097\\
N                   &     787,628&     628,022\\
\bottomrule
\end{tabular}
}
}
\caption*{\footnotesize Notes. We report semi-elasticities (see \cref{footnote:correction}), with standard errors clustered at the bi-directional market level in parentheses. Other control variables included in all regressions, but whose coefficients are not reported are \texttt{Talk-Eligible}, \texttt{Monopoly},  \texttt{MissingReport}, interactions of the \texttt{MissingReport} indicator with \texttt{Talk-Eligible} and \texttt{Monopoly}, origin- and destination-airport annual time trends, carrier-year-quarter fixed effects, and carrier-market fixed effects. All markets that include New York City, NY, or Washington D.C. are excluded in column 2.
}
\end{threeparttable} \end{table}

We use the same specification as \cref{eq:model_capdis}, except for the markets' city-pair definition. The primary results are in \cref{table:city.main}, column (1). The interpretations of all variables are the same as before, and the coefficient of interest is the first row, which shows that under this alternative approach to defining markets, communication does not correlate with the offered seats. 

To further shed light on why communication seems to be uncorrelated with capacity, we begin by observing that only two cities have three airports. 
Our result may be driven by what is happening in those two cities. So we re-estimate the model, but without Washington, D.C., (which includes BWI, DCA, and IAD) and New York City (EWR, JFK, and LGA), and present these results in column (2) of \cref{table:city.main}. As we can see, in city-pairs served by at most two airports, the capacity discipline parameter estimates are equal to -1.85\%, which is similar to the -2.02\% we found for the airport-pair markets. Thus, these two cities with three airports (Washington, D.C., and New York City) appear to be driving the differences between our primary, airport-pair market results and these city-pair market results. To understand the reason behind these differences, we need to understand the role of airports in the coordination mechanism, which is beyond our paper's scope.

\setcounter{table}{0}
\renewcommand{\thetable}{\Alph{section}.\arabic{table}}
\setcounter{figure}{0}
\renewcommand{\thefigure}{\Alph{section}.\arabic{figure}}

\section{Independent Verification}\label{appendix:ind_verification}
In \cref{section:ML} we detail the process we employ to code whether carriers discuss capacity discipline in each transcript. In this appendix, we consider two approaches to ensure that our results are not affected by the way we coded.

%\subsection*{RA Coding}

In the first approach, we hired an undergraduate student majoring in economics from the University of Virginia. We provided the student with our definition of ``capacity discipline,'' and then had the student read every transcript and independently decide whether an earnings call discussed capacity discipline. Similar to our approach in \cref{section:ML}, the student classified cases where the exact words ``capacity discipline" were used and where the exact words do not appear, but the concept of ``capacity discipline" was discussed. A detailed description of the RA's coding and the associated table is available upon request.

%\subsection*{NLP Coding}

In the second approach, we used natural language processing tools to automatically code each transcript based on whether a variation of the phrase ``capacity discipline'' was used. In this approach, we relied entirely on the automatic processing of the transcripts, rather than augmenting that work with human inspection of transcripts.

%\subsection*{Empirical Results}

\cref{tab:ra_results} shows the results of estimating our primary model under these two approaches. The first column shows the results of estimating this model using the RA's transcript coding data, and the second column shows the results of using the machine-coded transcripts. To aid in comparison, we reproduce our main results, from the first column of \cref{tab.main_results}, in the third column of \cref{tab:ra_results}. We find similar estimates to what we present in \cref{tab.main_results} under both the RA and Automatic coding approaches.

\begin{table}[t!] \centering
 \caption{Estimates from Independently Classified Data}\label{tab:ra_results}
\begin{threeparttable} \scalebox{0.9}{
{
\def\sym#1{\ifmmode^{#1}\else\(^{#1}\)\fi}
\begin{tabular}{l*{3}{c}}
\toprule
                    &\multicolumn{1}{c}{(1)}&\multicolumn{1}{c}{(2)}&\multicolumn{1}{c}{(3)}\\
                    &\multicolumn{1}{c}{Log Seats}&\multicolumn{1}{c}{Log Seats}&\multicolumn{1}{c}{Log Seats}\\
\midrule
Capacity Discipline &     -0.0215&     -0.0197&     -0.0202\\
                    &    (0.0077)&    (0.0071)&    (0.0060)\\
\midrule
R-squared           &       0.088&       0.088&       0.088\\
N                   &     841,991&     841,991&     841,991\\
\bottomrule
\end{tabular}
}

}

{\footnotesize Notes. In column 1, we present the results of estimating \cref{eq:model_capdis} using a communication variable that was independently coded by an RA, and in column 2 we use a communication variable that was automatically coded using natural language processing tools. Column 3 reports our primary estimates, \cref{tab.main_results}, column 1 to aid comparisons across these three approaches. We report semi-elasticities (see \cref{footnote:correction}), with standard errors clustered at the bi-directional market level in parentheses. Other control variables included in all regressions, but whose coefficients are not reported are \texttt{Talk-Eligible}, \texttt{Monopoly},  \texttt{MissingReport}, interactions of the \texttt{MissingReport} indicator with \texttt{Talk-Eligible} and \texttt{Monopoly}, origin- and destination-airport annual time trends, carrier-year-quarter fixed effects, and carrier-market fixed effects.
}
\end{threeparttable}
\end{table}

\section{Alternative Alignment of Earnings Call and Airline Data\label{section:realquarters}}

\begin{table}[ht!]
\begin{center}
\caption{Communication and Available Seats: Alternative Timing \label{tab:oldquarter}}
	\begin{adjustbox}{width=.9\textwidth}
		{
\def\sym#1{\ifmmode^{#1}\else\(^{#1}\)\fi}
\begin{tabular}{l*{3}{c}}
\toprule
                    &\multicolumn{1}{c}{(1)}&\multicolumn{1}{c}{(2)}&\multicolumn{1}{c}{(3)}\\
                    &\multicolumn{1}{c}{Log Seats}&\multicolumn{1}{c}{Log Seats}&\multicolumn{1}{c}{Log Seats}\\
\midrule
Capacity-Discipline &     -0.0210&            &            \\
                    &    (0.0058)&            &            \\
Capacity Discipline 2&            &     -0.0202&            \\
                    &            &    (0.0063)&            \\
Capacity Discipline 3&            &     -0.0278&            \\
                    &            &    (0.0106)&            \\
Capacity Discipline 4&            &     -0.0468&            \\
                    &            &    (0.0378)&            \\
Legacy Market x Capacity-Discipline&            &            &     -0.0194\\
                    &            &            &    (0.0075)\\
Mixed Market x Capacity Discipline (Legacy)&            &            &     -0.0182\\
                    &            &            &    (0.0117)\\
Mixed Market x Capacity Discipline (LCC)&            &            &     -0.0327\\
                    &            &            &    (0.0149)\\
\midrule
R-squared           &       0.088&       0.088&       0.089\\
N                   &     841,991&     841,991&     841,991\\
\bottomrule
\end{tabular}
}

	\end{adjustbox}
	\end{center}
	{\footnotesize Notes. This table replicates the primary estimates (columns 1, 3, and 5) from \cref{tab.main_results}, except we now associate, e.g., the Q1 call taking place in mid-April with the airline capacity data for April, May, and June. We report semi-elasticities (see \cref{footnote:correction}), with standard errors clustered at the bi-directional market level in parentheses. Other control variables included in all regressions, but whose coefficients are not reported are \texttt{Talk-Eligible}, \texttt{Monopoly},  \texttt{MissingReport}, interactions of the \texttt{MissingReport} indicator with \texttt{Talk-Eligible} and \texttt{Monopoly}. In column 2, these coefficients are allowed to vary based on the number of legacy carriers in the market (either 0 or 1, 2, 3, 4, or 5 legacy carriers). In column 3, these coefficients are allowed to vary across legacy and mixed markets, and within mixed markets for legacy carriers and LCCs. Additionally, all regressions include origin- and destination-airport annual time trends, carrier-year-quarter fixed effects, and carrier-market fixed effects.
	}
	
\end{table}

As discussed in \cref{sec:data_merge}, an airline's earnings call about a specific quarter takes place following the conclusion of that quarter, and throughout the paper, we associate the content of an earnings call with the three full months following the call. For example, we use the content from a Q1 call from mid-April to define $\texttt{Carrier-Capacity-Discipline}_{j,t}$ for May, June, and July.
Alternatively, we could associate the Q1 call, taking place in mid-April, with the capacity data for \emph{April}, May, and June. In Appendix \ref{tab:oldquarter} we reproduce all the results in \cref{tab.main_results} using this alternative definition. As we can see, the estimates are similar to those in \cref{tab.main_results}, suggesting that our findings are robust with respect to this definition. 

\setcounter{table}{0}
\renewcommand{\thetable}{\Alph{section}.\arabic{table}}
\setcounter{figure}{0}
\renewcommand{\thefigure}{\Alph{section}.\arabic{figure}}
 
\section{Communication and Capacity Responses and the DOJ Investigation}\label{section:20102015}
We investigate whether the airlines appear to have behaved differently before and after two key moments in the legal cases regarding capacity discipline. First, we investigate whether the estimates vary before and after January 2010, reportedly the earliest date in the records requests the DOJ sent to the airlines (c.f. \cref{section:legal}). To this end, we estimate \cref{eq:model_capdis}, allowing the role of \texttt{Capacity-Discipline} before January 2010 to be different from the one after January 2010. The estimates are presented in column (1) of \cref{tab.prepost}. We find that the difference in the estimates before and after January, 2010 is not statistically significant.

Next, we consider whether the estimated parameter estimates vary before and after the \emph{Washington Post} article reporting the DOJ investigation was published in July 2015. The DOJ investigation began at approximately the same time. As before, we estimate \cref{eq:model_capdis}, allowing the relationship before the \emph{Washington Post} article to be different from the one after the article was published. We present these results in column (2) of \cref{tab.prepost}. We find that the relationship we estimated in the paper between capacity choices and communication does not appear to persist beyond the July 2015 announcement of the DOJ investigation. That said, it is important to note that Washington Post announcement occurs close to the end of our sample (which ends at the close of the third quarter of 2016), and so we observe substantially fewer carrier-market-months in the post period of this particular analysis.
\begin{table}[t!]
\begin{center}
\caption{Communication and Available Seats: Before \& After Key Moments in DOJ Investigation\label{tab.prepost}}
	\begin{adjustbox}{width=0.7\textwidth}
		{
\def\sym#1{\ifmmode^{#1}\else\(^{#1}\)\fi}
\begin{tabular}{l*{2}{c}}
\toprule
                    &\multicolumn{1}{c}{(1)}&\multicolumn{1}{c}{(2)}\\
                    &\multicolumn{1}{c}{Log Seats}&\multicolumn{1}{c}{Log Seats}\\
\midrule
Pre-2010 Capacity Discipline&     -0.0206&            \\
                    &    (0.0094)&            \\
Post-2010 Capacity Discipline&     -0.0193&            \\
                    &    (0.0068)&            \\
Pre-WaPo Capacity Discipline&            &     -0.0164\\
                    &            &    (0.0058)\\
Post-WaPo Capacity Discipline&            &      0.0597\\
                    &            &    (0.0231)\\
\midrule
R-squared           &       0.088&       0.088\\
N                   &     841,991&     841,991\\
\bottomrule
\end{tabular}
}
	
	\end{adjustbox}
\end{center}	
{\footnotesize Notes. We report semi-elasticities (see \cref{footnote:correction}), with standard errors clustered at the bi-directional market level in parentheses. Other control variables included in all regressions, but whose coefficients are not reported are \texttt{Talk-Eligible}, \texttt{Monopoly},  \texttt{MissingReport}, interactions of the \texttt{MissingReport} indicator with \texttt{Talk-Eligible} and \texttt{Monopoly}. These coefficients are allowed to vary before and after 2010 or the publication of the article in The Washington Post. Additionally, all regressions include origin- and destination-airport annual time trends, carrier-year-quarter fixed effects, and carrier-market fixed effects.
}
	
\end{table}

\section{Examples of Contents in Earnings Calls}

In this section, we discuss the contents of the earnings calls pertaining to capacity discipline, which can shed light on what the airline executives generally say when they discuss capacity discipline. Airlines typically mention capacity discipline as part of a broader discussion of their capacity plans or broader strategic goals, but what is said depends on several factors that are carrier-specific, such as their networks of airports served, exposure to the fluctuations in fuel costs, expectations about future demand, contracts with regional carriers and their labor contracts. The following quotes provide some context for how the topic of capacity discipline appear in discussions of an airline's strategic goals.

\begin{quote}
	``\dots and while we still have a long way to go, we believe we are moving down the right track by continuing our capacity discipline while we strengthen our balance sheet and reinvest in key products, services, and in our fleet.''-- American, 2007 Q2
\end{quote}

\begin{quote}
	``To get there, we're focused on these key points: growing diversified revenues, treating our people well in a culture of positive employee relations, continuing our capacity discipline, keep our costs under control, running an airline customers worldwide prefer, deleveraging the business and limiting capital spending through investments with high IRRs.'' -- Delta, 2011 Q3
\end{quote}

In this example, we see the airline specifically relate capacity discipline to cancellation decisions.
\begin{quote}
	``And in addition, and kind of in line with our capacity discipline strategy, we're taking a lot more aggressive approach on kind of day/week cancellations, particularly in the sub UA network this Thanksgiving versus the past. '' -- United, 2011 Q3
\end{quote}

At times, airlines specifically note that they are comparing capacity growth to GDP growth.
\begin{quote}
	``As you can see, we remain committed to keeping our capacity discipline in check and our capacity growth within GDP rates.'' -- Delta, 2010 Q3
\end{quote}

From their conversations, we can also deduce that airlines understand that there are benefits from capacity discipline but that these benefits accrue only if their competitors also exercise capacity discipline. For instance, consider the following quotes from Alaska and United:
\begin{quote}
	``So we mentioned that Delta is trending upward in our markets. But we are actually seeing really good capacity discipline from other carriers on the West Coast, in particular from United, from Virgin, and from Southwest making pretty material reductions in our network.'' -- Alaska, 2014 Q1
\end{quote}

\begin{quote}
	``So again, I think our capacity discipline, as well as the industry discipline, what we've seen, I think, we've done a good job of not---the traffic that we're missing is the low yield price-sensitive traffic and we're doing a good job of not diluting the higher-end traffic. And I think the capacity discipline has allowed us to do that.'' -- United, 2011 Q3
\end{quote}

In other cases, we see airlines discuss capacity decisions in the context of their competitors' behavior, though without specifically raising the phrase ``capacity discipline.''
\begin{quote} 
	``We have taken steps to further trim our domestic capacity for 2003. But I think American on its own making small incremental reductions in capacity don't really help solve the overall industry imbalance between capacity and demand and just put us at a further competitive disadvantage\dots We also continue to plan for reduced capacity on a year-over-year basis. For the third quarter we expect mainline capacity to be down more than 5\% from last year's third quarter\dots Additionally, with a better alignment of capacity and demand this year the industry may well benefit from a reasonably stable pricing environment.'' -- AA, 2002 Q4
\end{quote}

As we can see in \cref{figure:capdis}, some airlines discuss capacity discipline less frequently than others. For instance, AS discusses capacity discipline less frequently than AA, but their ``messages" are similar whenever they discuss. The differences in what they say appear to be a function of the differences in the markets they serve---AS's business is mostly concentrated in the Northwest region, as exemplified below (slightly edited for clarity).
\begin{quote}
	``[W]hat you are referring to are reductions announced by Delta and Northwest. We have almost no overlap with them so their capacity reductions really don't help us. But you might hypothesize that a capacity reduction in other markets in the country might cause competitors' capacity to move to fill that void and that might moderate what we would have seen otherwise in competitive incursions in our geography. But I would say the impact that we expect from those capacity reductions on AS is very small. And we are not a player in the transcontinental except from Seattle and there hasn't been any big reduction capacity in the transcontinental from the Seattle market." -- AS, 2005 Q3
\end{quote}

During the period we study, several airlines file for bankruptcies, and we find that their competitors appear to keep track of their capacity plans. This concern is captured nicely in the following: 
\begin{quote}{``We pulled down a fair bit of capacity this summer. ...[O]ne of the questions for the whole industry is at significantly higher ticket prices, what does the demand picture look like and then how much excess capacity is there? It's exacerbated a little bit by the movement in competitor's capacity, ...while domestic capacity is down about 5\% in 2006, that's not what we're seeing within our geography. Within our geography we're seeing competitive capacity [up about] 3\%. But you know, we're hopeful, and we have got to see what happens to the rest of the capacity and how carriers [act with] bankruptcy for this year. We're sort of watching what's happening, with Independence Air going away, and with Delta and Northwest bankruptcy, their shrinking capacity in the Heartland and on the East Coast, and we're not big players in either of those markets... They're moving some capacity in the West Coast markets that they pulled during the bankruptcy, so we are a little bit concerned about that."
}--AS, 2005 Q4\end{quote}

In discussing capacity discipline, airlines also discuss various ways in which they might get rid of their ``excess" capacity. 
These methods include a mix of reducing capacity buying plan, re-writing contracts with their regional partners, expediting retirement of aircraft, delaying future aircrafts deliveries, re-allocating capacities to international markets, where the mix and thus the savings vary across airlines.

\setcounter{table}{0}
\renewcommand{\thetable}{\Alph{section}.\arabic{table}}
\setcounter{figure}{0}
\renewcommand{\thefigure}{\Alph{section}.\arabic{figure}}

\section{Additional Results}\label{section:additional_fe}
In this section, we present the results from estimating models in Tables \ref{table:robustness}, \ref{tab.pop_biz}, \ref{tab.WhiteChalak}, \ref{table:city.main}, \ref{tab:ra_results} and \ref{tab:oldquarter}, with carrier \textit{market-structure} fixed effects.

\begin{table}[t!] \centering \caption{Financial Transparency and Information Sharing}\label{table:robustness_new}
\begin{threeparttable} \scalebox{0.9}{
{
\def\sym#1{\ifmmode^{#1}\else\(^{#1}\)\fi}
\begin{tabular}{l*{5}{c}}
\toprule
                    &\multicolumn{1}{c}{(1)}&\multicolumn{1}{c}{(2)}&\multicolumn{1}{c}{(3)}&\multicolumn{1}{c}{(4)}&\multicolumn{1}{c}{(5)}\\
                    &\multicolumn{1}{c}{Log Seats}&\multicolumn{1}{c}{Log Seats}&\multicolumn{1}{c}{Log Seats}&\multicolumn{1}{c}{Log Seats}&\multicolumn{1}{c}{Log Seats}\\
\midrule
Only $j$ Talks      &      0.0136&            &            &            &            \\
                    &    (0.0054)&            &            &            &            \\
Monopoly Capacity Discipline&            &      0.0294&      0.0027&            &            \\
                    &            &    (0.0064)&    (0.0050)&            &            \\
Capacity Discipline $N-1$&            &            &            &     -0.0001&            \\
                    &            &            &            &    (0.0039)&            \\
Capacity Discipline ``Not $j$''&            &            &            &            &     -0.0026\\
                    &            &            &            &            &    (0.0060)\\
\midrule
R-squared           &       0.083&       0.084&       0.061&       0.083&       0.083\\
N                   &     841,991&     841,991&     438,980&     841,991&     841,991\\
\bottomrule
\end{tabular}
}

}
\caption*{\footnotesize Notes. We report semi-elasticities (see \cref{footnote:correction}), with standard errors clustered at the bi-directional market level in parentheses. Other control variables included in all regressions, but whose coefficients are not reported are \texttt{Talk-Eligible}, \texttt{Monopoly},  \texttt{MissingReport}, interactions of the \texttt{MissingReport} indicator with \texttt{Talk-Eligible} and \texttt{Monopoly}, origin- and destination-airport annual time trends, carrier-year-quarter fixed effects, and carrier-market-structure fixed effects. Column 3 omits the \texttt{Talk-Eligible} and \texttt{Monopoly} variables.
}

\end{threeparttable}
\end{table}

\begin{table}[t!] \centering \caption{Estimates for Conditional Exogeneity (Carrier-Market-Structure Fixed Effects)}\label{tab.WhiteChalak_new}
\begin{threeparttable} \scalebox{0.85}{
{
\def\sym#1{\ifmmode^{#1}\else\(^{#1}\)\fi}
\begin{tabular}{l*{6}{c}}
\toprule
                    &\multicolumn{1}{c}{(1)}&\multicolumn{1}{c}{(2)}&\multicolumn{1}{c}{(3)}&\multicolumn{1}{c}{(4)}&\multicolumn{1}{c}{(5)}&\multicolumn{1}{c}{(6)}\\
                    &\multicolumn{1}{c}{slow}&\multicolumn{1}{c}{weakness}&\multicolumn{1}{c}{domestically}&\multicolumn{1}{c}{internationally}&\multicolumn{1}{c}{stable}&\multicolumn{1}{c}{pace}\\
\midrule
Z Token             &     -0.0026&      0.0022&      0.0131&     -0.0008&     -0.0041&      0.0057\\
                    &    (0.0056)&    (0.0060)&    (0.0060)&    (0.0047)&    (0.0087)&    (0.0065)\\
Capacity Discipline &     -0.0148&     -0.0150&     -0.0144&     -0.0150&     -0.0148&     -0.0154\\
                    &    (0.0049)&    (0.0049)&    (0.0049)&    (0.0049)&    (0.0049)&    (0.0050)\\
\midrule
N                   &     841,991&     841,991&     841,991&     841,991&     841,991&     841,991\\
\bottomrule
\end{tabular}
}

}
\caption*{\footnotesize Notes. Estimation results from including new tokens an additional regressors in \cref{eq:model_capdis}. The table shows the coefficient estimates for each token, and for \texttt{Capacity-Discipline}. We report semi-elasticities (see \cref{footnote:correction}), with standard errors clustered at the bi-directional market level in parentheses. Other control variables included in all regressions, but whose coefficients are not reported are \texttt{Talk-Eligible}, \texttt{Monopoly},  \texttt{MissingReport}, interactions of the \texttt{MissingReport} indicator with \texttt{Talk-Eligible} and \texttt{Monopoly}, origin- and destination-airport annual time trends, carrier-year-quarter fixed effects, and carrier-market-structure fixed effects.
}
\end{threeparttable} \end{table}

\begin{table}[t!]\centering
\caption{Communication and Available Seats: The Role of Market Size and Business Travel}\label{tab.pop_biz_new}
\begin{threeparttable} \scalebox{0.75}{
{
\def\sym#1{\ifmmode^{#1}\else\(^{#1}\)\fi}
\begin{tabular}{l*{2}{c}}
\toprule
                    &\multicolumn{1}{c}{(1)}&\multicolumn{1}{c}{(2)}\\
                    &\multicolumn{1}{c}{Log Seats}&\multicolumn{1}{c}{Log Seats}\\
\midrule
Capacity Discipline x Small Population&     -0.0060&            \\
                    &    (0.0252)&            \\
Capacity Discipline x Medium Population&     -0.0083&            \\
                    &    (0.0091)&            \\
Capacity Discipline x Large Population&     -0.0189&            \\
                    &    (0.0054)&            \\
Capacity Discipline x Low Business&            &     -0.0131\\
                    &            &    (0.0080)\\
Capacity Discipline x Medium Business&            &     -0.0122\\
                    &            &    (0.0066)\\
Capacity Discipline x High Business&            &     -0.0233\\
                    &            &    (0.0090)\\
\midrule
R-squared           &       0.084&       0.082\\
N                   &     841,991&     620,762\\
\bottomrule
\end{tabular}
}

}
\caption*{\footnotesize Notes. We report semi-elasticities (see \cref{footnote:correction}), with standard errors clustered at the bi-directional market level in parentheses. Other control variables included in all regressions, but whose coefficients are not reported are \texttt{Talk-Eligible}, \texttt{Monopoly},  \texttt{MissingReport}, interactions of the \texttt{MissingReport} indicator with \texttt{Talk-Eligible} and \texttt{Monopoly}. These coefficients are allowed to vary based on the market size or business travel classifiers. Additionally, all regressions include origin- and destination-airport annual time trends, carrier-year-quarter fixed effects, and carrier-market-structure fixed effects.}
\end{threeparttable}
\end{table}

\begin{table}[t!] \centering \caption{Communication and Available Seats for City-Pair Markets (Carrier-Market-Structure Fixed Effects)} \begin{threeparttable} \scalebox{0.8}{
{
\def\sym#1{\ifmmode^{#1}\else\(^{#1}\)\fi}
\begin{tabular}{l*{2}{c}}
\toprule
                    &\multicolumn{1}{c}{(1)}&\multicolumn{1}{c}{(2)}\\
                    &\multicolumn{1}{c}{Log Seats}&\multicolumn{1}{c}{Log Seats}\\
\midrule
Capacity Discipline &     -0.0020&     -0.0088\\
                    &    (0.0040)&    (0.0050)\\
Exclude NYC \& DC & No & Yes\\\midrule
R-squared           &       0.086&       0.092\\
N                   &     787,628&     628,022\\
\bottomrule
\end{tabular}
}

}\label{table:city.main_new}
\caption*{\footnotesize Notes. We report semi-elasticities (see \cref{footnote:correction}), with standard errors clustered at the bi-directional market level in parentheses. Other control variables included in all regressions, but whose coefficients are not reported are \texttt{Talk-Eligible}, \texttt{Monopoly},  \texttt{MissingReport}, interactions of the \texttt{MissingReport} indicator with \texttt{Talk-Eligible} and \texttt{Monopoly}, origin- and destination-airport annual time trends, carrier-year-quarter fixed effects, and carrier-market-structure fixed effects. All markets that include New York City, NY, or Washington D.C. are excluded in column 2.
}
\end{threeparttable} \end{table}

\begin{table}[t!] \centering
 \caption{Estimates from Independently Classified Data (Carrier-Market-Structure Fixed Effects)}\label{tab:ra_results_new}
\begin{threeparttable} \scalebox{0.9}{
{
\def\sym#1{\ifmmode^{#1}\else\(^{#1}\)\fi}
\begin{tabular}{l*{3}{c}}
\toprule
                    &\multicolumn{1}{c}{(1)}&\multicolumn{1}{c}{(2)}&\multicolumn{1}{c}{(3)}\\
                    &\multicolumn{1}{c}{Log Seats}&\multicolumn{1}{c}{Log Seats}&\multicolumn{1}{c}{Log Seats}\\
\midrule
Capacity Discipline &     -0.0169&     -0.0181&     -0.0150\\
                    &    (0.0061)&    (0.0057)&    (0.0049)\\
\midrule
R-squared           &       0.083&       0.083&       0.083\\
N                   &     841,991&     841,991&     841,991\\
\bottomrule
\end{tabular}
}

}

{\footnotesize Notes. In column 1, we present the results of estimating \cref{eq:model_capdis} using a communication variable that was independently coded by an RA, and in column 2 we use a communication variable that was automatically coded using natural language processing tools. Column 3 reports our primary estimates, \cref{tab.main_results}, column 1 to aid comparisons across these three approaches. We report semi-elasticities (see \cref{footnote:correction}), with standard errors clustered at the bi-directional market level in parentheses. Other control variables included in all regressions, but whose coefficients are not reported are \texttt{Talk-Eligible}, \texttt{Monopoly},  \texttt{MissingReport}, interactions of the \texttt{MissingReport} indicator with \texttt{Talk-Eligible} and \texttt{Monopoly}, origin- and destination-airport annual time trends, carrier-year-quarter fixed effects, and carrier-market-structure fixed effects.
}
\end{threeparttable}
\end{table}

\begin{table}[ht!]
\begin{center}
\caption{Communication and Available Seats: Alternative Timing  (Carrier-Market-Structure Fixed Effects)\label{tab:oldquarter_new}}
	\begin{adjustbox}{width=.9\textwidth}
		{
\def\sym#1{\ifmmode^{#1}\else\(^{#1}\)\fi}
\begin{tabular}{l*{3}{c}}
\toprule
                    &\multicolumn{1}{c}{(1)}&\multicolumn{1}{c}{(2)}&\multicolumn{1}{c}{(3)}\\
                    &\multicolumn{1}{c}{Log Seats}&\multicolumn{1}{c}{Log Seats}&\multicolumn{1}{c}{Log Seats}\\
\midrule
Capacity-Discipline &     -0.0166&            &            \\
                    &    (0.0049)&            &            \\
Capacity Discipline 2&            &     -0.0171&            \\
                    &            &    (0.0052)&            \\
Capacity Discipline 3&            &     -0.0152&            \\
                    &            &    (0.0101)&            \\
Capacity Discipline 4&            &     -0.0286&            \\
                    &            &    (0.0221)&            \\
Legacy Market x Capacity-Discipline&            &            &     -0.0223\\
                    &            &            &    (0.0065)\\
Mixed Market x Capacity Discipline (Legacy)&            &            &      0.0003\\
                    &            &            &    (0.0086)\\
Mixed Market x Capacity Discipline (LCC)&            &            &     -0.0213\\
                    &            &            &    (0.0095)\\
\midrule
R-squared           &       0.084&       0.084&       0.084\\
N                   &     841,991&     841,991&     841,991\\
\bottomrule
\end{tabular}
}

	\end{adjustbox}
	
	{\footnotesize Notes. This table replicates the primary estimates (columns 1, 3, and 5) from \cref{tab.main_results}, except we now associate, e.g., the Q1 call taking place in mid-April with the airline capacity data for April, May, and June. We report semi-elasticities (see \cref{footnote:correction}), with standard errors clustered at the bi-directional market level in parentheses. Other control variables included in all regressions, but whose coefficients are not reported are \texttt{Talk-Eligible}, \texttt{Monopoly},  \texttt{MissingReport}, interactions of the \texttt{MissingReport} indicator with \texttt{Talk-Eligible} and \texttt{Monopoly}. In column 2, these coefficients are allowed to vary based on the number of legacy carriers in the market (either 0 or 1, 2, 3, 4, or 5 legacy carriers). In column 3, these coefficients are allowed to vary across legacy and mixed markets, and within mixed markets for legacy carriers and LCCs. Additionally, all regressions include origin- and destination-airport annual time trends, carrier-year-quarter fixed effects, and carrier-market-structure fixed effects.
	}
	\end{center}
\end{table}

\begin{table}[t!]
\begin{center}
\caption{Communication and Available Seats: Before \& After Key Moments in DOJ Investigation  (Carrier-Market-Structure Fixed Effects)\label{tab.prepost_new}}
	\begin{adjustbox}{width=0.7\textwidth}
		{
\def\sym#1{\ifmmode^{#1}\else\(^{#1}\)\fi}
\begin{tabular}{l*{2}{c}}
\toprule
                    &\multicolumn{1}{c}{(1)}&\multicolumn{1}{c}{(2)}\\
                    &\multicolumn{1}{c}{Log Seats}&\multicolumn{1}{c}{Log Seats}\\
\midrule
Pre-2010 Capacity Discipline&     -0.0229&            \\
                    &    (0.0081)&            \\
Post-2010 Capacity Discipline&     -0.0110&            \\
                    &    (0.0061)&            \\
Pre-WaPo Capacity Discipline&            &     -0.0157\\
                    &            &    (0.0050)\\
Post-WaPo Capacity Discipline&            &      0.0776\\
                    &            &    (0.0220)\\
\midrule
R-squared           &       0.084&       0.084\\
N                   &     841,991&     841,991\\
\bottomrule
\end{tabular}
}
	
	\end{adjustbox}
	
{\footnotesize Notes. We report semi-elasticities (see \cref{footnote:correction}), with standard errors clustered at the bi-directional market level in parentheses. Other control variables included in all regressions, but whose coefficients are not reported are \texttt{Talk-Eligible}, \texttt{Monopoly},  \texttt{MissingReport}, interactions of the \texttt{MissingReport} indicator with \texttt{Talk-Eligible} and \texttt{Monopoly}. These coefficients are allowed to vary before and after 2010 or the publication of the article in The Washington Post. Additionally, all regressions include origin- and destination-airport annual time trends, carrier-year-quarter fixed effects, and carrier-market-structure fixed effects.
}
	\end{center}
\end{table}

\clearpage
\renewcommand{\refname}{Appendix References}
\bibliographystyle{aea}
\bibliography{bibliography}

\end{appendices}
\end{document}